\documentclass[11pt]{article}
\usepackage[english]{babel}
\usepackage[latin1]{inputenc}
\usepackage{amsmath,dsfont}
\usepackage{color}
\usepackage{amsfonts}
\usepackage{graphicx} \usepackage{amsmath,amssymb}
\textheight by \topskip \textwidth 470pt
\makeatletter \@addtoreset{equation}{section}
\def\theequation{\thesection.\arabic{equation}}
\makeatother\def\Z{\mathbb Z}
\def\R{\mathbb R}

\def\be{\begin{equation}}
\def\ee{\end{equation}}\def\f{\frac}\def\sech{{\rm sech}}

\def\bea{\begin{eqnarray}}\def\eea{\end{eqnarray}}

\def\ben{\begin{displaymath}}
\def\ba{\begin{array}{c}}\def\bal{\begin{array}{l}}\def\ea{\end{array}}

\def\Z{\mathbb Z}
\def\A{\mathbb A}
\def\P{\mathbb P}

\def\R{\mathbb R}

\def\een{\end{displaymath}}

\begin{document}
\title{Effect of scalings and translations on
the supersymmetric quantum mechanical structure of soliton
systems}
\author{\textsf{Adri\'an Arancibia${}^{a}$, Juan Mateos Guilarte${}^{b}$ and
Mikhail S. Plyushchay${}^{a}$}
\\
[4pt]
 {\small \textit{${}^{a}$ Departamento de F\'{\i}sica, Universidad de
Santiago de Chile, Casilla 307, Santiago 2,
Chile}}\\
{\small \textit{${}^{b}$ Departamento de Fisica Fundamental
and IUFFyM, Universidad de Salamanca, Spain  }}\\
\sl{\small{E-mails: adaran.phi@gmail.com, guilarte@usal.es,
mikhail.plyushchay@usach.cl} }}
\date{}
\date{}

\maketitle
\begin{abstract}
We investigate a peculiar supersymmetry of the pairs of
reflectionless quantum mechanical systems described by $n$-soliton
potentials of a general form that depends on $n$ scaling and $n$
translation parameters. We show that if all the discrete energy
levels of the subsystems are different, the superalgebra, being
insensitive to translation parameters, is generated by two
supercharges of differential order $2n$, two supercharges of order
$2n+1$, and two bosonic integrals of order $2n+1$ composed from
Lax integrals of the partners. The exotic supersymmetry undergoes
a reduction when $r$ discrete energy levels of one subsystem
coincide with any $r$ discrete levels of the partner, the total
order of the two independent intertwining generators reduces then
to $4n-2r+1$, and the nonlinear superalgebraic structure acquires
a dependence on $r$ relative translations. For a complete pairwise
coincidence of the scaling parameters which control the energies
of the bound states and the transmission scattering amplitudes,
the emerging isospectrality is detected by a transmutation  of one
of the Lax integrals into a bosonic central charge. Within the
isospectral class, we reveal a special case giving a new family of
finite-gap first order Bogoliubov-de Gennes systems related to the
AKNS integrable hierarchy.
\end{abstract}

\section{Introduction}

Solitons and related topologically nontrivial objects such as kinks,
instantons, vortices, monopoles and domain walls play an important
role in diverse areas of physics, engineering and biology
\cite{Raj,DraJoh,ManSut}. Darboux and B\"acklund transformations,
with their origin in the theory of the linear Sturm-Liouville
problem and classical differential geometry,  proved to be very
effective in their study \cite{MatSal,RogSch}. Darboux
transformations \cite{MatSal}, on the other hand, underlie the
construction of supersymmetric quantum mechanics
\cite{Wit,CooKhaSuk}. Via  the Bogomolny bound and the associated
first order Bogomolny-Prasad-Sommerfield equations
\cite{Bogomolny,PraSom}, supersymmetry, in turn,  turns out to be
closely related with the topological solitons
\cite{WitOli,DvaShi,GibTow}.

Solitons and their periodic analogs appear as solutions of classical
nonlinear integrable field equations, and by means of Lax
representation \cite{Lax} are related with reflectionless and
periodic finite-gap quantum systems \cite{Belo,GesHol}. As both
families of quantum systems are characterized by nontrivial, higher
derivative integrals of motion, one could expect that supersymmetric
extensions of them should possess some peculiar properties. This is
indeed the case
\cite{BraMac,DunFei,FerNN,FerSam1,CorPly1,IofMatVal}, and exotic
supersymmetric structures of reflectionless and finite-gap systems
found recently some interesting physical applications
\cite{CJNP,CDP,PlyANie,JakPly,Jak}.

The most known example of reflectionless systems is given by a
hierarchy of P\"oschl-Teller potentials. The Schr\"odinger
Hamiltonian with one-, two-, and, in general, $n$ bound states
P\"oschl-Teller reflectionless potentials controls, particularly,
the stability of kinks in sine-Gordon, $\varphi^4$ and other exotic
(1+1)-dimensional field theory models
\cite{Raj,ManSut,Jack,TruFle,BoyCas,GraJaf,GolRebNieWim,AloMat1}.
These systems also appear in Gross-Neveu model \cite{DHN,Fei1}. The
indicated hierarchy represents, however, only a very restricted case
of a general family of $n$-soliton potentials. The latter
corresponds to $2n$-parametric solutions of the Korteweg-de Vries
(KdV) equation \cite{DraJoh,MatSal,PerZel}.

More explicitly, the Schr\"odinger operator is at the heart of the
inverse scattering transform method of solving the classical KdV
equation, for which the reflectionless potentials $V_n$ provide the
particle-like, $n$-soliton solutions. On the other hand, the
Schr\"odinger Hamiltonians $H=-\frac{d^2}{dx^2}+V_n$ with
reflectionless potentials $V_n$ control the stability of the above
mentioned kink solutions in $(1+1)$-dimensional field theories, and
their certain supersymmetric quantum mechanical structure proved
particularly to be very useful in the computing of the kink mass
quantum shifts, see ref. \cite{AJM}.

In the present paper we study the exotic supersymmetry that
appears in the pairs of reflectionless systems described by
$n$-soliton potentials of the most general form. Namely, we
investigate a peculiar supersymmetric quantum mechanical structure
of the class of one-dimensional systems described  by a matrix
$2\times 2$
 Hamiltonian
\begin{equation}\label{Hspin_n}
    \mathcal{H}=\left(%
\begin{array}{cc}
  -\frac{d^2}{dx^2}+V_+(x) & 0 \\
  0 & -\frac{d^2}{dx^2}+V_-(x) \\
\end{array}%
\right)\,,
\end{equation}
with
\begin{equation}\label{V+V-n}
    V_+(x)=V_n(x,\vec{\kappa},\vec{\tau})\quad
    \text{and}\quad
    V_-(x)=V_n(x,\vec{\kappa}\,',\vec{\tau}\,')
\end{equation}
 to be  $n$-soliton solutions of the KdV equation,
each depending on the sets of $n$ scaling parameters, denoted here
as  $\vec{\kappa}$ and $\vec{\kappa}\,'$, and $n$ translation
parameters, $\vec{\tau}$ and $\vec{\tau}\,'$. One of the possible
(but not unique, see below) physical interpretations of the system
(\ref{Hspin_n}), (\ref{V+V-n})  is that it can be considered as a
Hamiltonian of non-relativistic spin-$1/2$ particle with
spin-dependent forces of a special form (not inducing spin flips).

A non-soliton system of a general form (\ref{Hspin_n}), with
arbitrary chosen potentials $V_+(x)$ and $V_-(x)$, has just a
trivial integral given by the diagonal Pauli matrix $\sigma_3$.
For a special choice of potentials $V_\pm=W^2(x)\pm\frac{dW}{dx}$,
this trivial symmetry is extended for supersymmetric structure
related to nontrivial additional integrals of motion
$Q_1=-i\frac{d}{dx}\sigma_1+\sigma_2W(x)$, $Q_2=i\sigma_3Q_1$.
They generate a linear in $\mathcal{H}$,  Lie superalgebraic
structure $\{Q_a,Q_b\}=2\delta_{ab}\mathcal{H}$,
$[\mathcal{H},Q_a]=0$, $a,b=1,2$, with the integral $\sigma_3$
playing a role of the $\Z_2$-grading operator,
$[\sigma_3,\mathcal{H}]=0$, $\{\sigma_3,Q_a\}=0$. It is such a
linear superalgebraic structure that appears, particularly, in the
Landau problem for non-relativistic electron, where superpotential
is a linear function $W(x)=\omega x$, and (\ref{Hspin_n}) takes a
form of the superoscillator Hamiltonian, see \cite{CooKhaSuk}. The
existence of the linear supersymmetric structure is equivalent to
the condition that the upper and lower components of the matrix
Hamiltonian, $H_\pm=-\frac{d^2}{dx^2}+V_\pm$, are related by the
Darboux intertwining generators, $H_+A_+=A_+H_-$, $H_-A_-=A_-H_+$,
being the first order differential operators
$A_+=\frac{d}{dx}+W(x)$ and $A_-=A_+^\dagger=-\frac{d}{dx}+W(x)$.
With this observation, the construction can be generalized to
nonlinear supersymmetry if the potentials $V_+$ and $V_-$ are such
that the corresponding partner Hamiltonians are connected by the
intertwining  relations of the same form, but with $A_+$ and
$A_-=A_+^\dagger$ to be differentials operators of order $\ell>1$.
If this happens, the system $\mathcal{H}$ possesses nilpotent
supercharges $Q_+=A_+\sigma_+=\frac{1}{2}(Q_2+iQ_1)$ and
$Q_-=A_-\sigma_-=Q_+^\dagger$,  $[Q_\pm, H]=0$, $Q_\pm^2=0$, where
$\sigma_\pm=\frac{1}{2}(\sigma_1\pm i\sigma_2)$. They generate a
nonlinear supersymmetry of the form
$\{Q_a,Q_b\}=2\delta_{ab}P_\ell(\mathcal{H})$, where
$P_\ell(\mathcal{H})$ is an order $\ell$ polynomial. The simplest
example of a system with nonlinear supersymmetry is provided by a
generalized superoscillator system $\mathcal{H}=b^+b^-+
\ell\frac{1}{2}(1+\sigma_3)$, for which $A_+=(b^-)^\ell$, $b^\pm$
are the usual creation-annihilation bosonic oscillator operators,
and the order $\ell$ polynomial is
$P_\ell(\mathcal{H})=\prod_{j=0}^{\ell-1}(H-j \omega)$, see ref.
\cite{PlKlanom}.

The peculiarity of the system (\ref{Hspin_n}), (\ref{V+V-n}) we
study here is that the $n$-soliton potentials (\ref{V+V-n}) are
reflectionless. By a known construction based on Crum-Darboux
transformations, such potentials can be obtained from a free
particle system, which possesses a momentum integral
$p=-i\frac{d}{dx}$. It will be shown that,  as a consequence, the
$n$-soliton extended system is described by  an exotic
supersymmetric structure that includes not only one but two pairs
of $Z_2$-odd (anti-diagonal) matrix supercharges,  and two
$Z_2$-even (diagonal) additional nontrivial bosonic integrals
being differential operators  of order $2n+1$. The supercharges in
general case are higher order matrix differential operators, two
of which are of the even order $2r$, and other two supercharges
are of the odd order $2l+1$ such that $2(r+l)\geq 2n$.
Corresponding superalgebra generated by four supercharges is
nonlinear, and includes in its structure those additional
nontrivial bosonic integrals of motion which are nothing else as a
Crum-Darboux dressed form of the free particle momentum operator.
The supercharges also have a nature of the dressed integrals of
motion of the free spin-$1/2$ particle described by the
Hamiltonian (\ref{Hspin_n}) with $V_+=V_-=0$. We shall show that
such a peculiar supersymmetric structure of the extended
$n$-soliton systems experiences radical changes in dependence on
relation between the two sets of the scaling and translation
parameters of the partner potentials: the differential order of
supercharges can change, and in the completely isospectral case
when $\vec{\kappa}=\vec{\kappa}\,'$, one of the additional bosonic
integrals transforms into the central charge of the corresponding
nonlinear superalgebra. Analyzing different faces of supersymmetry
restructuring,  we detect, particularly,  a special family of
supersymmetric $n$-soliton partner potentials when one pair of
supercharges reduces to the matrix first order differential
operators. These first order supercharges and $\mathcal{H}$ form
between themselves a linear superalgebra corresponding to the
broken supersymmetry. In such a case, one of the first order
supercharges can be reinterpreted as a first order Hamiltonian of
a Dirac particle. The reinterpretation provides us then with new
kink-anti-kink type solutions for the Gross-Neveu model by means
of the first order Bogoliubov-de Gennes system, in which a
superpotential takes a meaning of a condensate, an order
parameter, or a gap function depending on the physical context.
 \vskip0.1cm

The paper is organized as follows. In the next Section, we review
the general construction of soliton potentials with the help of
Crum-Darboux transformations, summarize the basic properties of
the corresponding reflectionless quantum systems, and  formulate
precisely the problems related to supersymmetry of  soliton
systems (\ref{Hspin_n}), (\ref{V+V-n}) to be studied here. Section
3 is devoted to the analysis  of supersymmetry of non-isospectral
pairs of reflectionless $n=1$  systems with different bound state
energy levels given in terms of non-equal scaling parameters
$\kappa_1\neq \kappa'_1$. In Section 4 we investigate the changes
this supersymmetric structure undergoes in the isospectral case
$\kappa_1= \kappa'_1$. Section 5 generalizes the results of
Section 3 for the case of $n>1$-soliton pairs with completely
broken isospectrality. To clarify the supersymmetry picture in
extended $n>1$ systems with partially broken and exact
isospectralities,  we study in detail the case of $n=2$ in Section
6. In Section 6.1 we review the properties of the generic $n=2$
reflectionless systems to identify the ingredients to be important
for further analysis. Then, in Section 6.2, we discuss a
generalization of Crum-Darboux transformations that is related to
alternative factorizations of the basic Crum-Darboux generators of
order $n>1$. The results of Sections 6.1 and  6.2  are employed in
Sections 6.3 and 6.4 for analysis of supersymmetry in extended
$n=2$ systems with partial isospectrality breaking. Finally, in
Sections 6.5, 6.6 and 6.7 we investigate the most tricky case of
supersymmetry in two-soliton extended  systems with exact
isospectrality. We do this first in Section 6.5 for a particular
case of exact isospectrality with a common virtual $n=1$
subsystem. In Section 6.6 we investigate a generic case of exact
isospectrality, within which we detect yet another, very special,
particular case. The latter is studied in Section 6.7, and
provides us with a new, first order finite-gap system belonging to
the AKNS hierarchy \cite{AKNS,GesHol}. In Section 7 we discuss how
the results on partially broken and exact isospectralities are
generalized for the systems (\ref{Hspin_n}), (\ref{V+V-n}) with
$n>2$.
In Section 8 we consider an interpretation of the system
(\ref{Hspin_n}), (\ref{V+V-n}) as a non-relativistic spin-$1/2$
particle with spin-dependent forces. We conclude the paper with
discussion of the obtained results and their possible developments
and applications in Section~9.

\section{Family of reflectionless $n$-soliton
systems}\label{SecCD}

A Crum-Darboux transformation  of order $n$, $n=1,2,\ldots$, applied
to a quantum free particle generates a  system characterized by the
Hamiltonian \cite{MatSal}
\begin{equation}
    H_n=H_0+V_n(x),\qquad
    V_n=-2\frac{d^2}{dx^2}\ln W_n\,. \label{lnW}
\end{equation}
Here $H_0=-\frac{d^2}{dx^2}$ is a free particle Hamiltonian, and
$W_n=W(\psi_1,\ldots,\psi_n)$ is a Wronskian of its  eigenfunctions
$\psi_1(x)$, $\ldots$, $\psi_n(x)$, $H_0\psi_j=E_j^{(0)}\psi_j$,
\begin{equation}\label{Wron-n}
    W(f_1,\ldots,f_n)=\det \mathcal{A},\qquad
    \mathcal{A}_{ij}=\frac{d^{i-1}}{dx^{i-1}}f_j,\qquad
    i,j=1,\ldots, n.
\end{equation}
A simple choice of $\psi_j(x)$ in the form of the unidirectional
plane waves  $e^{ik_jx}$, which are eingensolutions of $H_0$,
produces the Wronskian of the form $W_n(x)=const\cdot
e^{i(k_1+\ldots + k_n)x}$, and, therefore, $V_n=0$. If we take a
linear independent set of linear combinations of left- and right-
moving  plane waves $\psi_j(x)=e^{ik_jx}+c_je^{-ik_jx}$ with
$c_j\neq 0$ for all $j=1,\ldots,n$, we obtain a nontrivial potential
$V_n\not\equiv 0$, which  satisfies a higher order stationary
$g$-KdV, $g=2n+1$, (Novikov) equation being a nonlinear ordinary
differential equation with a linear highest derivative
${d\,^{g}V_n}/{dx^{g}}$ term \cite{Novikov,GesWei}. (\ref{lnW})
belongs then to a class of finite-gap, or algebro-geometric systems
\footnote{Finite-gap \emph{periodic} systems are given by the
Its-Matveev representation of the form (\ref{lnW}) but with $W(x)$
substituted by a Riemann's theta function \cite{ItsMat}. If such a
periodic potential is real and regular on $\R$, the spectrum of
Schr\"odinger (Hill) operator is organized in valence and a
conductance bands separated by gaps. (\ref{lnW}) with
reflectionless, $n$-soliton potential (\ref{VnBarg}) can be
considered then as the infinite period limit of a periodic or almost
periodic finite-gap system. In the indicated limit, the valence
bands shrink, some of which can merge in this process, and transform
into the non-degenerate discrete energy levels of the bound states
of a resulting soliton potential; the semi-infinite conductance band
turns into the continuous part of the spectrum of a reflectionless
system. Quantum systems with periodic $n$-gap and non-periodic
$n$-soliton potentials (whose discrete energy levels and continuous
spectrum are also separated by $n$ gaps) are characterized by the
existence of the differential operator of order $2n+1$, related with
a higher order Novikov equation, that commutes with a Hamiltonian,
see below. A free particle can be treated in this picture as a
zero-gap system (of an arbitrary period), for which the
corresponding first order differential operator is just the momentum
integral $p=-i\frac{d}{dx}$. For the theory of finite-gap and
soliton systems including historical aspects, see
\cite{Belo,FinRev}.}. For real $k_j$, the emergent `finite-gap'
potential $V_n(x)$ has, however, singularities on $\R$ and does not
disappear at $x=\pm \infty$. An appropriate choice of the free
particle non-physical eigenfunctions (corresponding to certain
linear combinations of the left- and right- moving plane waves
evaluated at imaginary momenta),
\begin{equation}\label{psij}
    \psi_j=\left\{\begin{matrix}\cosh \kappa_j(x+\tau_j),&
    j={\rm odd}\\ \sinh
    \kappa_j(x+\tau_j),&j={\rm even}\end{matrix}\,\,,\right.
    \qquad 0<\kappa_1<\kappa_2<...<\kappa_{j-1}<\kappa_n\,,\quad
\end{equation}
of energies $E_j^{(0)}=-\kappa^2_j$, $j=1,\ldots,n,$ gives rise to
a nodeless Wronskian $W_n(x)$.  A non-singular $2n$-parametric
potential
\begin{equation}\label{VnBarg}
    V_n=V_n(x;\kappa_1,\tau_1,\ldots,\kappa_n,\tau_n)
\end{equation}
corresponds then to a \emph{reflectionless} (Bargmann) system $H_n$
with $n+1$ non-degenerate states, separated by $n$ gaps, $n$ of
which, of energies $E^{(n)}_j=-\kappa^2_j$, $j=1,\ldots, n$, are the
bound states, while the non-degenerate state of zero energy, $E=0$,
lies at the bottom of the doubly degenerate continuous spectrum with
$E>0$. From another perspective, reflectionless potential
$V_n(x;\kappa_1,\tau_1,\ldots,\kappa_n,\tau_n)$ describes
$n$-soliton solutions of the KdV equation.

Eigenstates of $H_n$, $H_n\psi[n;\lambda]=\lambda\psi[n;\lambda]$,
different from the physical bound states, are generated from
eigenfunctions $\psi[0;\lambda]$ of the free particle,
$H_0\psi[0;\lambda]=\lambda\psi[0;\lambda]$, $\lambda\neq
-\kappa_j^2$,
\begin{equation}\label{psilam}
    \psi[n;\lambda]=\frac{W(\psi_1,\ldots,\psi_n,
    \psi[0;\lambda])}{W(\psi_1,\ldots,\psi_n)}\,,
\end{equation}
where $\psi_j$ are given by Eq. (\ref{psij}). Physical
non-degenerate bound states of $H_n$ with $\lambda=-\kappa_j^2$ are
obtained by the same prescription (\ref{psilam}) under the choice
$\psi[0;\lambda]= \sinh \kappa_j(x+\tau_j)$ for odd $j$, and
$\psi[0;\lambda]= \cosh \kappa_j(x+\tau_j)$ for even $j$. The lowest
non-degenerate state of the continuous part of the spectrum of $H_n$
corresponds to the eigenstate $\psi[0;0]=1$ of $H_0$.

Transmission scattering amplitudes $a[n;k]$ for the continuous part
of the spectrum $E=k^2$, $k>0$, of reflectionless system $H_n$ are
defined by the scaling parameters $\kappa_j$ \cite{MatSal},
\begin{equation}\label{trans}
    a[n;k]=\prod_{j=1}^{n}\frac{k-i\kappa_j}{k+i\kappa_j}\,.
\end{equation}

The states (\ref{psilam}) have an alternative but equivalent
representation, $\psi[n;\lambda]=A_n\ldots A_1\psi[0;\lambda]$,
generated by an $n$-sequence of the first order Darboux
transformations,
\begin{equation}\label{Darbj}
    \psi[j;\lambda]\equiv
    \psi[(\kappa,\tau)_{(j)};\lambda]=
    A_j\psi[j-1;\lambda]\,,
\end{equation}
where $(\kappa,\tau)_{(j)}$ denotes the set of $2j$ parameters
$\kappa_1,\tau_1,\ldots,\kappa_j,\tau_j$, and
$A_j=A_j[(\kappa,\tau)_{(j)}]$ are the first order differential
operators defined recursively in terms of the states (\ref{psij}) by
\begin{equation}\label{A1def}
    A_1=\psi_1 \frac{d}{dx} \frac{1}{\psi_1}=
    \frac{d}{dx} -\kappa_1\tanh\kappa_1(x+\tau_1)\,,
\end{equation}
\begin{equation}\label{Ajdef}
    A_j=(A_{j-1}\ldots A_1\psi_j)\frac{d}{dx} \frac{1}{(A_{j-1}\ldots
    A_1\psi_j)}=\frac{d}{dx} -\left(\frac{d}{dx} \ln(A_{j-1}\ldots
    A_1\psi_j)\right)\,.
\end{equation}
The first order operator $A_j$ annihilates the state $A_{j-1}\ldots
A_1\psi_j$, that is a nonphysical eigenstate of $H_{j-1}$ of
eigenvalue $-\kappa_{j}^2$. As inverse  to (\ref{Darbj}), there is,
up to an overall multiplicative constant, a relation
\begin{equation}\label{Darbj+}
    \psi[j-1;\lambda]=
    A_j^\dagger \psi[j;\lambda]\,.
\end{equation}
The zero mode of the first order operator $A_j^\dagger$ is
$1/(A_{j-1}\ldots A_1\psi_j)$. It is the ground state of $H_j$ of
the energy $-\kappa_j^2$.

A reflectionless $j$-soliton Hamiltonian $H_j$ admits two
factorization representations
\begin{equation}\label{Hjfactor}
    H_j=A_{j+1}^\dagger A_{j+1}-\kappa_{j+1}^2=
    A_{j} A_{j}^\dagger -\kappa_{j}^2\,.
\end{equation}
In particular, the free particle $0$-gap Hamiltonian
$H_0=-\frac{d^2}{dx^2}$ has an alternative representation
$
    H_0=A_{1}^\dagger A_{1}-\kappa_{1}^2.
$
{}From (\ref{Hjfactor}) there follow intertwining relations
\begin{equation}\label{interj}
    A_jH_{j-1}=H_jA_j\,,\qquad
    A_j^\dagger H_j=H_{j-1}A_j^\dagger\,,\qquad
    j=1,\ldots, n\,.
\end{equation}

Let us take now a pair of $n$-soliton reflectionless systems,
\begin{equation}\label{2Hn}
    H_n=H_n(\kappa_1,\tau_1,\ldots,\kappa_n,\tau_n)
    \quad {\rm and}\quad
    H_n'=H_n(\kappa_1',\tau_1',\ldots,\kappa_n',\tau_n')\,,
\end{equation}
and consider the extended matrix $2\times 2$ Hamiltonian of the
form (\ref{Hspin_n}) with $H_+=H_n$ and $H_-=H_n'$. Two sets of
parameters are supposed to be completely different, or may
partially coincide. If the two sets of the scaling parameters
$\kappa_j$, $j=1,\ldots,n$, and $\kappa'_{j'}$, $j'=1,\ldots,n$,
do not coincide, the two subsystems have not only different
spectra of bound states, but in accordance with (\ref{trans}),
their transmission amplitudes are also different. If, moreover,
$\kappa_j\neq \kappa'_{j'}$ for all $j,j'=1,\ldots, n$, all the
energy levels of bound states for two $n$-soliton reflectionless
systems are different, and  their transmission amplitudes are
given by rational functions of $k$ with different zeroes and
poles. Having in mind that the factorization relations
(\ref{Hjfactor}) and the associated intertwining relations
(\ref{interj}) are reformulated in terms of supersymmetric quantum
mechanics construction, one can put a question:
 \begin{itemize}
\item
    What a  supersymmetric structure is
    associated with reflectionless pair (\ref{2Hn}) in  a
    \emph{completely non-isospectral} case\footnote{
    Using this term we neglect the fact that the continuous
    (scattering) parts of the
    spectra of the partner systems are the same, $E\geq 0$.} characterized by inequalities
    $\kappa_j\neq \kappa'_{j'}$ for all $j,j'=1,\ldots, n$?
\end{itemize}
Such a kind of supersymmetry of the pairs of reflectionless
systems was not investigated yet in the literature, but, instead,
supersymmetry of the pairs ($H_+=H_j$, $H_-=H_{j+l}$), $l\geq 1$,
belonging to the same Darboux chain (\ref{interj}) is usually
considered. In particular, the pairs of reflectionless
P\"oschl-Teller systems, see below, appear in the context of
shape-invariance \cite{Gend,CooGinKha,CooKhaSuk}, they also emerge
in the infinite-period limit of finite-gap periodic crystal
structures \cite{CJNP,PlyANie}. Supersymmetry of reflectionless
P\"oschl-Teller pairs ($H_j$, $H_{j+l}$) was studied recently from
the perspective of AdS/CFT holography and Aharonov-Bohm effect
\cite{AdSPT}.

A special choice of the parameteres
\begin{equation}\label{ktPT}
    \kappa_j=\kappa_j'=j\kappa,
    \qquad \tau_j=\tau,\qquad
    \tau_j'=\tau',\quad j=1,\ldots, n\,,
\end{equation}
 results in two copies of the
$n$-soliton potentials $V_n=-n(n+1)\kappa^2\sech^2\kappa(x+\tau)$
and $V_n'=-n(n+1)\kappa^2\sech^2\kappa(x+\tau')$, which describe two
mutually shifted  reflectionless P\"oschl-Teller systems with $n$
bound states. Since the partner potentials under the choice
(\ref{ktPT}) have exactly the same form,  this corresponds to a
particular case of a shape-invariance, whose analog in the case of
periodic supersymmetric systems was called by Dunne and Feinberg
`\emph{self-isospectrality}' \cite{DunFei}. The exotic nonlinear
supersymmetry of the simplest isospectral pair $(H_+=H_1$,
$H_-=H_1')$ with $\kappa_1=\kappa_1'$, $\tau_1\neq \tau_1'$ was
investigated and applied for the description of the kink and
kink-anti-kink solutions of the Gross-Neveu model
\cite{PlyNie,PlyANie}. One can expect that the self-isospectral pair
of  reflectionless P\"oschl-Teller systems with $n>1$ bound states
should also be described by some not studied yet exotic nonlinear
supersymmetric structure.

In a more general case of the choice $\kappa_j=\kappa_j'$,
$j=1,\ldots,n$,  different from (\ref{ktPT}), the partners  with
$\vec{\tau}\neq\vec{\tau}\,'$, $\vec{\tau}=(\tau_1,\ldots, \tau_n)$,
are completely isospectral, their bound states energies and
transmission amplitudes coincide, but the potentials have different
form. We then arrive at the natural questions related to that
formulated above:
\begin{itemize}
\item
How the supersymmetric structure  of a general,  non-isospectral
case detects the coincidence of some of the scaling parameters of
two systems in (\ref{2Hn})?
\item
Particularly,  for a partial coincidence of the bound states energy
levels, does the supersymmetry distinguish the coincidence of the
scaling parameters of the same level, $\kappa_j=\kappa'_j$, from
that corresponding to the case when distinct levels,
$\kappa_j=\kappa'_{j'}$ with $j\neq j'$, coincide?
\item
Is the case of a complete isospectrality of the two systems,
$\kappa_j=\kappa'_j$, $j=1,\ldots, n$, detected somehow by
supersymmetric structure?
\item
Does the case  of self-isospectrality possess some special
characteristics from the viewpoint of supersymmetry in comparison
with a general case of isospectral systems with different form of
potentials?
\end{itemize}

In what follows,  we study a  peculiar supersymmetric structure of
the pair (\ref{2Hn}), and, particularly, respond the highlighted
questions.

\section{Supersymmetry of $n=1$
reflectionless pair with distinct scalings}\label{n1non}

We first investigate the supersymmetric structure of the extended
system
\begin{equation}\label{H1ext}
    \mathcal{H}_1=\left(
    \begin{array}{cc}
      H_1 & 0 \\
      0 & H_1' \\
    \end{array}
\right)
\end{equation}
described by the pair of $n=1$ reflectionless P\"oschl-Teller
Hamiltonians $H_1=H_1(\kappa,\tau)$  and $H_1'=H_1(\kappa',\tau')$
with $\kappa\neq \kappa'$ and arbitrary displacement parameters
$\tau$ and $\tau'$. This will allow us to trace how the
restructuring of supersymmetry happens in the self-isospectral case
$\kappa=\kappa'$, and to form a base for further analysis for $n>1$,
where we will restore index $1$, omitted here to simplify notations,
in the scaling and translation parameters.

The choice of a non-physical eigenstate
$\psi_1(\kappa,\tau)=\cosh{\kappa(x+\tau)}$, $\kappa>0$,  $\tau \in
\R$, of $H_0$ produces a Hamiltonian of $n=1$ reflectionless
P\"oschl-Teller system
\begin{equation}\label{h1}
H_1=-\frac{d^2}{dx^2}-\f{2\kappa^2}{\cosh^2\kappa (x+\tau)}\,,
\end{equation}
and first order operators  $A_1$  and $A_1^\dagger$ defined by Eq.
(\ref{A1def}). Operators $A_1$ and $A_1^\dagger$ factorize the
shifted for an additive constant Hamiltonians $H_0$ and $H_1$,
\begin{eqnarray}\label{AAH01}
    H_1=A_1
    A_1^\dag-
    \kappa^2,&\quad& H_{0}=
    A_1^\dag A_1-\kappa^2\,,
\end{eqnarray}
and intertwine them,
\begin{equation}\label{H0H1int}
    A_1^\dag H_1=
    H_{0} A_1^\dag\,,\qquad
    A_1H_{0}=
    H_1A_1\,.
\end{equation}

A degenerate pair of eigenstates in the continuous part, $E=k^2$,
$k>0$, of the spectrum of $H_1$ is constructed from the free
particle plane wave states,
\begin{equation}\label{H1cont}
    \psi^{\pm k}_{1}=
    A_1(\kappa,\tau)e^{\pm ikx}=(\pm ik-
    \kappa\tanh{\kappa (x+\tau)})e^{\pm ikx}\,.
\end{equation}
The lowest non-degenerate state  with $E=0$ corresponds to a
boundary case $k=0$ of (\ref{H1cont}),
\begin{equation}
    \psi^{0}_{1}=\tanh\kappa(x+\tau)\,.
\end{equation}
Another,  bound non-degenerate state
\begin{equation}\label{psiligado}
    \psi^{-\kappa^2}_1=
    \kappa\,{\rm sech}\, \kappa (x+\tau)
\end{equation}
of energy $E=-\kappa^2$ is obtained from the  partner,
$\tilde{\psi}_1(\kappa,\tau)=\sinh \kappa(x+\tau)$, of non-physical
eigenstate $\psi_1(\kappa,\tau)=\cosh{\kappa(x+\tau)}$ of $H_0$, $
\psi^{-\kappa^2}_1(\kappa,\tau)=A_1(\kappa,\tau)
\tilde{\psi}_1(\kappa,\tau)$.

Based on intertwining relations (\ref{H0H1int}) and their analog for
the system $H_1'=H_1(\kappa',\tau')$, we construct the second order
operator
\begin{equation}\label{Y1}
    Y_2=Y_2(\kappa,\tau,\kappa',\tau')=
    A_1(\kappa,\tau)A_1^\dag(\kappa',\tau')=A_1{A_1'}^\dag\,,\qquad
    Y_2^\dagger=
    Y_2(\kappa',\tau',\kappa,\tau)=Y_2'\,,
\end{equation}
that intertwines the partner Hamiltonians of the extended  system
(\ref{H1ext}), $    Y_2H_1'=
    H_1Y_2$.
 Taking into account
that $H_0$ has an integral $p=-i\frac{d}{dx}$, one can obtain yet
another, third order intertwining operator,
\begin{equation}\label{X1}
    X_3=X_3(\kappa,\tau,\kappa',\tau')=
    A_1\frac{d}{dx}{A_1'}^\dag\,,\qquad
    X_3^\dagger(\kappa,\tau,\kappa',\tau')=
     -X_3(\kappa',\tau',\kappa,\tau)=-X_3'\,,
\end{equation}
$X_3H_1'=
    H_1X_3$, which is independent from the second order intertwiner $Y_2$.

\begin{figure}[h!]\begin{center}
\includegraphics[scale=1.2]{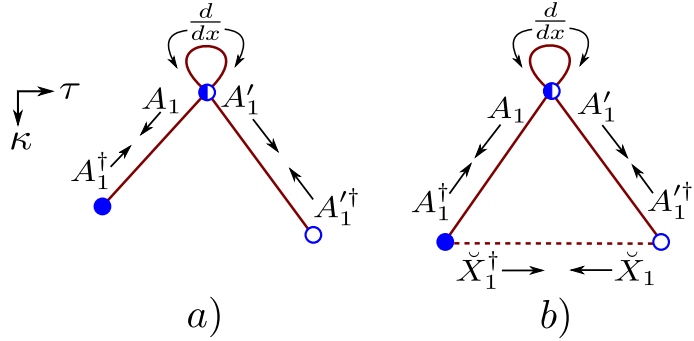}
\caption{a) Non-isospectral one-soliton  Hamiltonians  $H_1$ (blue
dot) and $H'_1$ (white dot) are intertwined by the second, $Y_2$ and
$Y_2^\dagger$, and the third, $X_3$ and $X_3^\dagger$, order
Crum-Darboux operators via a virtual translation-invariant free
particle system $H_0$ (half blue/half white dot). b) In the
isospectral case $\kappa=\kappa'$, a direct `tunneling' channel for
intertwining by the first order operators $\breve{X}_1$ and
$\breve{X}_1^\dagger$ is opened. In both cases, Lax integrals $Z_3$
and $Z'_3$, being the dressed forms of the free particle integral
$\frac{d}{dx}$, are the `self-intertwining' generators for $H_1$ and
$H_1'$.}\label{fig1}
\end{center}
\end{figure}

Intertwining  relations in the reverse direction are obtained by a
change $\kappa,\,\tau\leftrightarrow\kappa',\tau'$, that corresponds
to a Hermitian conjugation of  the corresponding relations,
$Y_2^\dagger H_1=H_1'Y_2^\dagger$, $X_3^\dagger
H_1=H_1'X_3^\dagger$, see Fig. \ref{fig1}a.

The free particle integral $p=-i\frac{d}{dx}$ and intertwining
relations (\ref{H0H1int}) also generate  a nontrivial integral for
the $n=1$ reflectionless P\"oschl-Teller subsystem
$H_1(\kappa,\tau)$,
\begin{equation}\label{P1}
    Z_3=Z_3(\kappa,\tau)=
    A_1\frac{d}{dx}A_1^\dagger\,,\qquad
    Z_3^\dagger=-Z_3\,,
\end{equation}
and the analogous integral, $Z_3'=A'_1\frac{d}{dx}A'^\dagger_1$, for
$H_1(\kappa',\tau')$. Integral (\ref{P1}) is a nontrivial operator
of a Lax pair for stationary KdV equation in the non-periodic case.

Here and in what follows, the odd and even order intertwining
operators are denoted   by $X$ and $Y$, respectively,  while the odd
order integrals of the corresponding reflectionless systems are
denoted by $Z$; the lower index indicates the differential order of
these operators.

Integral (\ref{P1}) detects both physical non-degenerate states of
$H_1(\kappa,\tau)$ by annihilating them $Z_3\psi_1^0(\kappa,\tau)=
Z_3\psi^{-\kappa^2}_1(\kappa,\tau)=0$. The third state of its kernel
is a non-physical eigenstate
$\tilde{\psi}^{-\kappa^2}_1(x)=\psi_1^{-\kappa^2}(x)\int
dx/(\psi_1^{-\kappa^2}(x))^2$ of $H_1$ of energy $-\kappa^2$, which
is a linear combination of the physical bound state
$\psi^{-\kappa^2}_1(x)$ of the same energy and of a non-physical
eigenstate $\psi_1(\kappa,\tau)=\cosh\kappa(x+\tau)$ of $H_0$.

The extended system (\ref{H1ext}) has an obvious integral of
motion $\sigma_3$.
 The intertwining relations together with integral (\ref{P1})
allow us to identify the nontrivial Hermitian integrals for the
system $\mathcal{H}_1$,
\begin{equation}\label{Q1int}
    \mathcal{Q}_{1;1}=\begin{pmatrix}0&Y_2\\
    Y_2^\dagger
    &0\end{pmatrix},\quad
    \mathcal{Q}_{1;2}=
    i\sigma_3\mathcal{Q}_{1;1},
    \qquad
    \mathcal{S}_{1;1}=
    \begin{pmatrix}0&X_3\\
    X_3^\dagger
    &0\end{pmatrix},\quad
    \mathcal{S}_{1;2}=
    i\sigma_3\mathcal{S}_{1;1},
\end{equation}
\begin{equation}\label{P1int}
    \mathcal{P}_{1;1}
    =-i
    \left(%
    \begin{array}{cc}
      Z_3& 0 \\
      0 & Z_3'\\
    \end{array}%
    \right),\qquad
     \mathcal{P}_{1;2}=
     \sigma_3\mathcal{P}_{1;1}\,.
\end{equation}
As $\sigma_3^2=\mathds{1}$, we can take the integral
$\Gamma=\sigma_3$ as a $\Z_2$-grading operator. It classifies then
$\mathcal{P}_{1;a}$, $a=1,2$, as bosonic integrals,
$[\sigma_3,\mathcal{P}_{1;a}]=0$, while the integrals (\ref{Q1int})
 are identified as fermionic supercharges,
 $\{\sigma_3,\mathcal{Q}_{1;a}\}=
 \{\sigma_3,\mathcal{S}_{1;a}\}=0
 $,
 of the supersymmetric
 structure of the extended system
 $\mathcal{H}_1$.
 There are other possibilities to choose
 $\Gamma$,
 which are based on  reflection operators and
 classify the nontrivial integrals
 of the extended system in a way different from that prescribed by the
 choice $\Gamma=\sigma_3$. The alternative
 choices for $\Gamma$ find some interesting physical applications,
 see \cite{CJNP,PlyANie,PlyNie,CJPjpa}, and we return to the discussion of
 this point in the last Section.

Operators (\ref{Q1int}) and (\ref{P1int}) are the Darboux-dressed
integrals  of the extended system described by the Hamiltonian
$\mathcal{H}_0={\rm diag}(H_0,H_0)$ composed from two copies of the
free particle Hamiltonian $H_0$. The system $\mathcal{H}_0$
possesses the set of  $2\times 2$ matrix Hermitian integrals
\begin{equation}\label{I0}
    \mathcal{I}_{0}=\sigma_a,\,\,\, \epsilon_{ab}\sigma_b p,\,\,\,
    \mathds{1}p,\,\,\, \sigma_3 p,
    \qquad a=1,2\,.
\end{equation}
The Darboux dressing,
 \begin{equation}\label{Idressed}
    \mathcal{I}_1=\mathcal{D}_1\mathcal{I}_0\mathcal{D}^\dagger_1\,,
    \qquad
    \mathcal{D}_1={\rm
    diag}\,(A_1(\kappa,\tau),A_1(\kappa',\tau'))\,,
\end{equation}
transforms them into the integrals (\ref{Q1int})  and (\ref{P1int})
of $\mathcal{H}_1$.

We find the superalgebraic structure of the system $\mathcal{H}_1$
by employing the intertwining and factorization relations
(\ref{H0H1int}) and (\ref{AAH01}). It is given by the following
nontrivial (anti)-commutation relations:
\begin{equation}\label{S1S1}
    \{\mathcal{Q}_{a},\mathcal{Q}_{b}\}=2\delta_{ab}
    \P_1(\mathcal{H}_1,\kappa) \P_1(\mathcal{H}_1,{\kappa}')
    \,,\quad
    \{\mathcal{S}_{a},\mathcal{S}_{b}\}=2\delta_{ab}
    \mathcal{H}_1\P_1(\mathcal{H}_1,\kappa)
    \P_1(\mathcal{H}_1,{\kappa}')\,,
\end{equation}
\begin{equation}\label{Q1S}
    \{\mathcal{S}_{a},\mathcal{Q}_{b}\}=2\epsilon_{ab}
    \P_1(\mathcal{H}_1,\mathcal{K})\mathcal{P}_{1}\,,
\end{equation}
\begin{equation}\label{P11QS}
     [\mathcal{P}_{1},\mathcal{S}_{a}]=i\mathcal{H}_1
     \P_0^-(\mathcal{H}_1,
    {\kappa},{\kappa}')\mathcal{Q}_{a}
    \,,\qquad
      [\mathcal{P}_{1},\mathcal{Q}_{a}]=-i\P_0^-(\mathcal{H}_1,
    {\kappa},{\kappa}')\mathcal{S}_{a}\,,
\end{equation}
\begin{equation}\label{P12QS}
     [\mathcal{P}_{2},\mathcal{S}_{a}]=i\mathcal{H}_1
     \P_1^+(\mathcal{H}_1,
    {\kappa},{\kappa}')\mathcal{Q}_{a}
    \,,\qquad
   [\mathcal{P}_{2},\mathcal{Q}_{a}]=-i\P_1^+(\mathcal{H}_1,
    {\kappa},{\kappa}')\mathcal{S}_{a}
    \,,
\end{equation}
where $\P_1(\mathcal{H}_1,{\kappa})=
    \mathcal{H}_1+
    \kappa^2\cdot\mathds{1},$
$\P_1(\mathcal{H}_1,\mathcal{K})=
    \mathcal{H}_1+
    \mathcal{K}^2,$
$\mathcal{K}=\text{diag}\, (\kappa',\kappa)$,
\begin{equation}\label{P-1}
    \P_0^-(\mathcal{H}_1,
    {\kappa},{\kappa}')=
    \P_1(\mathcal{H}_1,{\kappa})-
    \P_1(\mathcal{H}_1,{\kappa}')=(\kappa^2-\kappa'^2)
    \cdot\mathds{1}\,,
\end{equation}
 $\P_1^+(\mathcal{H}_1,
{\kappa},{\kappa}')= \P_1(\mathcal{H}_1,{\kappa})+
\P_1(\mathcal{H}_1,{\kappa}')=2\mathcal{H}_1+ (\kappa^2+\kappa'^2)
\cdot\mathds{1}$, and to simplify the formulae, we omitted the index
$n=1$ in the supercharges and bosonic integrals. Though in the final
expression for $\P_0^-$ in (\ref{P-1}) the dependence on
$\mathcal{H}_1$ disappears, it is indicated here in the arguments
having in mind a further generalization for the $n>1$ case, where
this structure is substituted for the polynomial of order $n-1$ in
Hamiltonian.

The $n=1$ extended reflectionless system (\ref{H1ext}) is described
therefore by a nonlinear superalgebra generated by four fermionic
supercharges, $\mathcal{Q}_{1;a}$ and $\mathcal{S}_{1;a}$, and by
two bosonic integrals~\footnote{There are four bosonic integrals if
one counts the integrals $\mathcal{H}_1$ and $\sigma_3$.},
$\mathcal{P}_{1;a}$. The fermionic integrals are constructed from
the intertwining operators of the second and third orders, whose
composition produces nontrivial third order integrals of Lax pairs
of the $n=1$  non-isospectral subsystems. In this supersymmetric
structure, Hamiltonian plays a role of the multiplicative central
charge. The nonlinear superalgebra depends here on the scaling
parameters $\kappa$ and $\kappa'$ via the polynomials $\P_1$,
$\P_1^+$ and $\P_0^-$, but does not depend on the displacement
parameters $\tau$ and $\tau'$.

\section{Supersymmetry of the $n=1$ self-isospectral
pair}\label{1selfiso}

For the isospectral extended system $\mathcal{H}_1$ with
$\kappa=\kappa'$, the partner potentials have the same form and are
mutually displaced. This $n=1$ self-isospectral case is special from
the viewpoint of supersymmetric structure. As follows from
(\ref{P11QS}) and (\ref{P-1}), for $\kappa=\kappa'$ the integral
$\mathcal{P}_{1;1}$, composed from the third order integrals of Lax
pairs of superpartner subsystems, commutes with all the integrals,
and so, transmutes  into a bosonic central charge of the nonlinear
superalgebra. We show now that the supersymmetric structure in this
case  undergoes even more radical changes.

For $\kappa=\kappa'$ the following reduction takes
place~\footnote{A reduction of the third order intertwining
generators was discussed in a general form in \cite{IofNis},
however, with giving no special attention to a peculiar
supersymmetric structure we study here; see also \cite{AraPly}.} :
\begin{equation}\label{Xbr}
    X_3(\kappa,\tau,\kappa,\tau')=\left(H_1(\kappa,\tau)+
    \kappa^2\right)\breve{X}_{1}(\kappa,\tau,\tau')-
    \mathcal{C}(\kappa,\tau-\tau')
    Y_2(\kappa,\tau,\kappa,\tau')\,,
\end{equation}
where
\begin{eqnarray}\label{X1breve}
     \breve{X}_{1}(\kappa,\tau,\tau')&=&
     \frac{d}{dx}-\kappa\tanh \kappa(x+\tau)+
     \kappa\tanh\kappa(x+\tau') +\mathcal{C}(\kappa,\tau-\tau')\\
     &=&
     A_1(\kappa,\tau)-A_1^\dagger(\kappa,\tau')+
     A_{\mathcal{C}}^\dagger(\kappa,\tau-\tau')
     \,,\label{AAC}
\end{eqnarray}
\begin{equation}\label{defC}
    A_{\mathcal{C}}(\kappa,\tau-\tau')=\frac{d}{dx}+\mathcal{C}(\kappa,\tau-\tau')\,,
    \qquad
    \mathcal{C}(\kappa,\tau-\tau')=\kappa
    \coth\kappa(\tau-\tau')\,.
\end{equation}

 Relation (\ref{Xbr}) means that for $\tau\neq\tau'$, the first
order operator $\breve{X}_{1}=\breve{X}_{1}(\kappa,\tau,\tau')$
should be taken as a basic odd order intertwining operator instead
of  $X_3(\kappa,\tau,\kappa,\tau')$,
\begin{equation}\label{X1brint}
    \breve{X}_{1}H_1(\kappa,\tau')=
    H_1(\kappa,\tau)\breve{X}_{1}\,,\qquad
    \breve{X}_{1}^\dagger(\kappa,\tau,\tau')=
    -\breve{X}_{1}(\kappa,\tau',\tau)=-\breve{X}_{1}'\,.
\end{equation}
Note that in the limit $\tau'\rightarrow\pm\infty$, we have
$H'_1\rightarrow H_0$ and $\breve{X}_1\rightarrow A_1$, while for
$\tau\rightarrow\pm\infty$,   $H_1\rightarrow H_0$ and
$\breve{X}_1\rightarrow -{A_1'}^\dagger$. This is coherent with the
intertwining relations (\ref{H0H1int}).

Because of  (\ref{Xbr}),  the third order integrals
$\mathcal{S}_{1;a}$  are reducible,
 $
    \mathcal{S}_{1;a}=(\mathcal{H}_1+\kappa^2)
    \breve{\mathcal{S}}_{1;a}-
    \mathcal{C}\mathcal{Q}_{1;a}\,,
$ and have to be changed for the first order irreducible integrals
\begin{equation}\label{S1brint}
    \breve{\mathcal{S}}_{1;1}=
    \begin{pmatrix}0&\breve{X}_1\\
    \breve{X}_1^\dagger
    &0\end{pmatrix},\qquad
    \breve{\mathcal{S}}_{1;2}=
    i\sigma_3\breve{\mathcal{S}}_{1;1}\,.
\end{equation}

Integrals $\breve{\mathcal{S}}_{1;a}$ correspond, in accordance with
(\ref{Idressed}), to the dressed form of the integrals
$\breve{s}_a=\epsilon_{ab}\sigma_bp +\mathcal{C}\sigma_a$,
 of the extended free particle system $\mathcal{H}_0={\rm
diag}(H_0,H_0)$, $\mathcal{D}\breve{s}_a\mathcal{D}^\dagger=
\breve{\mathcal{S}}_{1;a}(\mathcal{H}_1+\kappa^2)$. Alternatively,
the first order matrix operator  $\breve{s}_1=\sigma_2
p+\mathcal{C}\sigma_1$, or $\breve{s}_2=i\sigma_3\breve{s}_1$,  can
be considered as a first order Hamiltonian of the free Dirac
particle of mass $|\mathcal{C}|$ in (1+1) dimensions, while its
dressed form, $\breve{\mathcal{S}}_{1;1}$, can be identified as a
Bogoliubov-de Gennes Hamiltonian describing the kink-antikink
solution in the Gross-Neveu model \cite{DHN}. Function
$\Delta(\xi,\lambda)=\kappa\left(\tanh{(\xi-\lambda)}-
    \tanh{(\xi+\lambda)}+
    \coth{2\lambda} \right)$, that appears in the structure of
$\breve{X}_1$  with
    $\xi=\kappa(x+\frac{\tau+\tau'}{2})$ and
    $\lambda=-\kappa
     \frac{\tau-\tau'}{2}$, has then a sense of a gap function \cite{CDP}.

The following relations are valid:
\begin{equation}\label{XXbrH}
     \breve{X}_{1}\breve{X}_{1}^\dagger=H_1(\kappa,\tau)+
     \mathcal{C}^2\,,
\end{equation}
\begin{eqnarray}\label{XbrA}
     &\breve{X}_{1}
     A_1(\kappa,\tau')=A_1(\kappa,\tau)
     A_{\mathcal{C}}(\kappa,\tau-\tau')\,,\qquad
    A^\dag_1(\kappa,\tau)\breve{X}_{1}=
    A_{\mathcal{C}}(\kappa,\tau-\tau')
    A^\dag_1(\kappa,\tau')\,.&
\end{eqnarray}
The employment of (\ref{XXbrH}), (\ref{XbrA}) together with
(\ref{X1brint}) gives nontrivial nonlinear superalgebraic relations
\begin{equation}\label{SSbreve}
    \{\breve{\mathcal{S}}_{1;a},
    \breve{\mathcal{S}}_{1;b}\}=2\delta_{ab}
    h_{\mathcal{C}}\,,\qquad
    \{\mathcal{Q}_{1;a},\mathcal{Q}_{1;b}\}=2\delta_{ab}
    h^2_\kappa\,,
\end{equation}
\begin{equation}\label{QSbreve}
    \{\breve{\mathcal{S}}_{1;a},\mathcal{Q}_{1;b}\}=
    2\delta_{ab}\mathcal{C}
    h_\kappa+2\epsilon_{ab}
    \mathcal{P}_{1;1}\,,
\end{equation}
\begin{equation}\label{PSQbreve}
     [\mathcal{P}_{1;2},\breve{\mathcal{S}}_{1;a}]=2i
    (h_{\mathcal{C}}\mathcal{Q}_{1;a}-
    \mathcal{C}h_\kappa\breve{\mathcal{S}}_{1;a}
    )\,,\qquad
    [\mathcal{P}_{1;2},\mathcal{Q}_{1;a}]=
    2ih_\kappa(
    \mathcal{C}\mathcal{Q}_{1;a}-h_\kappa
    \breve{\mathcal{S}}_{1;a})\,,
\end{equation}
which substitute nontrivial superalgebraic relations (\ref{S1S1}),
(\ref{Q1S}), (\ref{P11QS}) and (\ref{P12QS}) of the general,
non-isospectral case $n=1$. Here we denoted
$h_\kappa=\mathcal{H}_1+\kappa^2$,
$h_{\mathcal{C}}=\mathcal{H}_1+\mathcal{C}^2$. As
$\mathcal{C}^2>\kappa^2$, the spectrum of $h_{\mathcal{C}}$ is
strictly positive, and the Lie sub-superalgebra generated by the
first order supercharges $\breve{\mathcal{S}}_{1;a}$ corresponds to
a broken $N=2$ supersymmetry. The $\mathcal{P}_{1;1}$ commutes now
with all the supercharges in accordance with the observation made at
the beginning of the Section.

While the third order intertwining operator (\ref{X1})  is well
defined at $\kappa=\kappa'$, $\tau=\tau'$ and reduces to the
integral $Z_3(\kappa,\tau)$ of $H_1(\kappa,\tau)$, the first order
intertwining operator $\breve{X}_{1}(\kappa,\tau,\tau')$ in the
limit $\tau'\rightarrow\tau$ reduces to the operator $\frac{d}{dx}$
shifted for an infinite additive constant term $\pm\infty$ in
dependence on which side the difference $(\tau-\tau')$ tends to
zero. In this case extended Hamiltonian (\ref{H1ext}) reduces just
to the two identical copies of the P\"oschl-Teller Hamiltonians,
$\mathcal{H}_1(\kappa,\tau)={\rm
diag}(H_1(\kappa,\tau),H_1(\kappa,\tau))$. The integrals
$\breve{\mathcal{S}}_{1;a}$, $a=1,2$,  can be renormalized
multiplying them by $1/\mathcal{C}(\kappa,\tau-\tau')$, and taking a
limit $\tau'\rightarrow \tau$. In such a way they are reduced to the
trivial integrals $\sigma_a$, $a=1,2$, of
$\mathcal{H}_1(\kappa,\tau)$. The second order intertwining operator
(\ref{Y1}) reduces in the limit $\tau'\rightarrow \tau$ to
$H_1(\kappa,\tau)+\kappa^2$, and the second order supercharges
$\mathcal{Q}_{1;a}$ are reduced to the same trivial integrals
$\sigma_a$ multiplied by a shifted for a constant Hamiltonian,
$Q_{1;a}\rightarrow(\mathcal{H}_1(\kappa,\tau)+\kappa^2\cdot
\mathds{1})\sigma_a$. The only nontrivial integrals we have  in the
limit $\tau'\rightarrow \tau$ are  the bosonic third order integrals
$\mathcal{P}_{1;a}(\kappa,\tau)$.

The special case of self-isospectrality in the $n=1$ extended system
$\mathcal{H}_1$ is detected, therefore, by a radical change of
nonlinear supersymmetric structure. One of the bosonic integrals,
$\mathcal{P}_{1;1}$, turns into a central charge, and two third
order supercharges are substituted for the supercharges of the first
order. The reduction of the order of the half of the supercharges at
$\kappa=\kappa'$  originates from relation (\ref{Xbr}) and is
accompanied by appearance of dependence of the superalgebraic
structure  on the distance between mutually shifted one-soliton
partner potentials by means of a constant $\mathcal{C}=\kappa\coth
\kappa(\tau-\tau')$. In other words, one can say that in a generic
case $\kappa\neq \kappa'$, the $H_1$ and $H_1'$ are intertwined by
the third order operators $X_3$ and $X_3^\dagger$, side by side with
the second order operators $Y_2$ and $Y_2^\dagger$, via the free
particle (zero gap) system, and the supersymmetric structure does
not feel a relative distance $\tau-\tau'$ between the corresponding
one-soliton subsystems because of the translation invariance of
$H_0$. For $\kappa=\kappa'$, a kind of a `tunneling' channel is
opened: the one-soliton subsystems are intertwined then directly by
the first order operators $\breve{X}_1$ and $\breve{X}_1^\dagger$,
and the modified supersymmetric structure detects a `tunneling
distance' $\tau-\tau'$, see Fig. \ref{fig1}b.

\section{Supersymmetry of an $n>1$ extended  system:
complete isospectrality breaking}\label{nisobrokencomp}

The discussion of supersymmetric structure for extended system
composed from two subsystems having $n\geq 2$ bound states requires
to distinguish three cases:
\begin{itemize}
\item Complete isospectrality breaking, when  $\kappa_i\neq
\kappa'_j$ for all $i,j=1,\ldots, n$, with no restriction on
displacement parameters $\tau_i$ and $\tau'_j$.
\item Partial
isospectrality breaking, in which some, but not all, scaling
parameters $\kappa_i$ and $\kappa'_j$ of the two subsystems
coincide. \item Exact isospectrality, that is characterized by the
complete coincidence of the sets of the scaling parameters,
$\vec{\kappa}=\vec{\kappa}'$, accompanied by a restriction
$\vec{\tau}\neq \vec{\tau}\,'$.
\end{itemize}

The case of a complete isospectrality breaking for $n>1$ is a direct
generalization of that for $n=1$ case with $\kappa_1\neq\kappa'_1$,
which was studied  in Section \ref{n1non}. It is discussed in the
present Section. Other two cases are more involved. Though they
generalize somehow the picture of the one-soliton case ($n=1$) with
$\kappa_1=\kappa_1'=\kappa$, investigated in the previous Section,
the corresponding analysis for $n>1$ requires a generalization of
the described Crum-Darboux transformations scheme. New peculiarities
appear there, and  those two cases deserve a separate consideration.
To understand the picture,  we study  the case of $n=2$  in the next
Section, and then in Section \ref{nisospectrality} the results will
be extended for a generic case of $n\geq 2$. \vskip0.2cm

\begin{figure}[h!]\begin{center}
\includegraphics[scale=1.2]{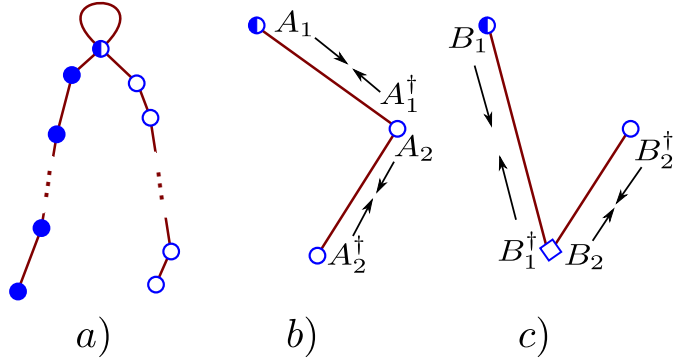}
\caption{a) An $n>1$ pair with complete isospectrality breaking.
Each subsystem,  $H_n$ and $H'_n$, is specified by indicating the
set of intermediate, virtual, systems in the plane $\kappa-\tau$ via
which the edge points are connected to the free particle  by means
of the first order Darboux generators $A_j$ and $A_j^\dagger$, not
shown here. Figs. b) and c) illustrate two alternative
representations for the same $n=2$ system, that is related to the
two different factorizations of the second order Crum-Darboux
generator $\A_2$. In the case b) the virtual system is regular,
while in the case c) it is singular. So, a system is specified not
only by indication of the set of points in the $\kappa-\tau$ plane,
but also by the path via these points to a free system
$H_0$.}\label{fig2}
\end{center}
\end{figure}

With these comments in mind, let us consider an extended system
\begin{equation}\label{Hnsusy}
    \mathcal{H}_n=\left(%
\begin{array}{cc}
  H_n  & 0 \\
  0 & H_n' \\
\end{array}%
\right),
\end{equation}
composed from a completely non-isospectral pair $H_n=
H_n(\vec{\kappa},\vec{\tau})$  and $H_n'=
H_n(\vec{\kappa}\,{}',\vec{\tau}\,{}')$ of the form (\ref{2Hn}),
where $\vec{\kappa}=(\kappa_1,\ldots,\kappa_n)$,
$\vec{\tau}=(\tau_1,\ldots,\tau_n)$, and  we assume that there is no
coincidence in the sets of the scaling parameters of the two
subsystems, $\kappa_j\neq \kappa'_{j'}$ for all $j,j'=1,\ldots, n$,
see Fig. \ref{fig2}a.

Following the general picture described in Section \ref{SecCD},
Hamiltonian $H_n=H_n(\vec{\kappa},\vec{\tau})$ can be intertwined
with a free particle Hamiltonian $H_0$ by order $n$ differential
operators $\A_n=\A_n(\vec{\kappa},\vec{\tau})$ and
$\A_n^\dagger=\A_n^\dagger(\vec{\kappa},\vec{\tau})$,
\begin{equation}\label{AnCD}
    \A_n(\vec{\kappa},\vec{\tau})=A_n((\kappa,\tau)_n)A_{n-1}((\kappa,\tau)_{n-1})\ldots
    A_1\left(\kappa_1,\tau_1\right)\,,
\end{equation}
defined in terms of Darboux generators (\ref{A1def}), (\ref{Ajdef}),
\begin{equation}\label{HnAn}
    \A_nH_0=
    H_n\A_n,\qquad
    \A_n^\dagger
    H_n=H_0\A_n^\dagger\,.
\end{equation}
 Making use of these
relations, we construct an order $2n$ operator
\begin{equation}\label{Yndef}
    Y_{2n}=Y_{2n}(\vec{\kappa},\vec{\tau};\vec{\kappa}\,',\vec{\tau}\,')=
    \A_n
    {\A_n'}^\dagger\,,\qquad
     Y_{2n}^\dagger=
      Y_{2n}(\vec{\kappa}\,',\vec{\tau}\,';\vec{\kappa},\vec{\tau})=Y_{2n}'\,,
\end{equation}
where ${\A_n'}=\A_n(\vec{\kappa}\,',\vec{\tau}\,')$, and two
operators of the order $2n+1$, $X_{2n+1}$ and $Z_{2n+1}$,
\begin{equation}\label{Xndef}
    X_{2n+1}(\vec{\kappa},\vec{\tau};\vec{\kappa}\,',\vec{\tau}\,')=
    \A_n\frac{d}{dx}
    {\A_n'}^\dagger\,,\qquad
     X_{2n+1}^\dagger=-
      X_{2n+1}(\vec{\kappa}\,',
      \vec{\tau}\,';\vec{\kappa},\vec{\tau})=-X_{2n+1}'\,,
\end{equation}
\begin{equation}\label{Pndef}
    Z_{2n+1}=Z_{2n+1}(\vec{\kappa},\vec{\tau})=
    \A_n\frac{d}{dx}
    \A_n^\dagger\,,\qquad
    Z_{2n+1}^\dagger=-Z_{2n+1}\,.
\end{equation}
Operators $Y_{2n}$ and $X_{2n+1}$ intertwine the components of the
matrix Hamiltonian $\mathcal{H}_n$~\footnote{Intertwining
relations through multi-step ladders of linear Darboux generators
and their superalgebraic reducibility have been recently reviewed
in \cite{AndIof}, but in a very general and abstract form.},
\begin{equation}\label{YnHn}
    Y_{2n}
    H_n'=
    H_n
    Y_{2n}\,,\qquad
    X_{2n+1}
    H_n'=
    H_n
    X_{2n+1}\,,
\end{equation}
while $Z_{2n+1}(\vec{\kappa},\vec{\tau})$ is an integral for
$H_n(\vec{\kappa},\vec{\tau})$,
\begin{equation}\label{PnHn}
    [Z_{2n+1},
    H_n]=0\,.
\end{equation}
Taking into account that the coefficients of the $(2n+1)$ order
differential operator  $Z_{2n+1}$ may be expressed in terms of the
potential $V_n$ and its derivatives of the order less than $2n+1$
\cite{GesWei}, relation (\ref{PnHn}) means that the potential $V_n$
satisfies a higher stationary $g$-KdV equation with $g=2n+1$,
mentioned in Section \ref{SecCD}.

In correspondence with an identity
$Z_{2n+1}^2=-H_n\prod_{j=1}^{j=n}(H_n+\kappa_j^2)^2$, the integral
$Z_{2n+1}$ detects all the physical non-degenerate states of $H_n$
of energies $E=0$ and $E_j=-\kappa_j^2$ by annihilating them. These
are constructed from the free particle non-degenerate eigenstate
$\psi_0^0=1$, $\psi_n^0=\A_n 1$,  and non-physical partners of the
states (\ref{psij}), $\tilde{\psi}_1=\sinh \kappa_1(x+\tau_1),$
$\tilde{\psi}_2=\cosh \kappa_2(x+\tau_2),\ldots,$
$\psi_n^{-\kappa_j^2}=\A_n\tilde{\psi}_j$, $j=1,\ldots, n$. Other
$n$ states of the kernel of $Z_{2n+1}$ are non-physical  partners of
the bound states $\psi_n^{-\kappa_j^2}$,
$\tilde{\psi}_n^{-\kappa_j^2}(x)=\psi_n^{-\kappa_j^2}(x)\int
dx/(\psi_n^{-\kappa_j^2}(x))^2$.

With the described operators, we construct six matrix integrals
$\mathcal{Q}_{n;a}$, $\mathcal{S}_{n;a}$ and $\mathcal{P}_{n;a}$ for
the extended system $\mathcal{H}_n$ in the form similar to that in
(\ref{Q1int}) and (\ref{P1int}) by changing $Y_2$, $X_3$ and $Z_3$
for, respectively, $Y_{2n}$, $X_{2n+1}$ and $Z_{2n+1}$. As in the
$n=1$ case, these integrals correspond to a dressed form of the
integrals of the extended free particle system $\mathcal{H}_0$
obtained by means of Eq. (\ref{Idressed}) with the change of
$\mathcal{D}_1$ for $\mathcal{D}_n={\rm
    diag}\,(\A(\vec{\kappa},\vec{\tau}),
    \A_n(\vec{\kappa}\,',\vec{\tau}\,'))$.

Applying factorization and intertwining relations, and products of
corresponding generators collected in Appendix, we find that the
superalgebra (\ref{S1S1}), (\ref{Q1S}), (\ref{P11QS}), (\ref{P12QS})
of the $n=1$ case is generalized for
\begin{equation}\label{QnQn}
    \{\mathcal{Q}_{n;a},\mathcal{Q}_{n;b}\}=2\delta_{ab}
    \P_n(\mathcal{H}_n,\vec{\kappa})
    \P_n(\mathcal{H}_n,\vec{\kappa}')\,,\,\,
    \{\mathcal{S}_{n;a},\mathcal{S}_{n;b}\}=2\delta_{ab}
    \mathcal{H}_n\P_n(\mathcal{H}_n,\vec{\kappa})
    \P_n(\mathcal{H}_n,\vec{\kappa}')\,,
\end{equation}
\begin{equation}\label{QnS}
    \{\mathcal{S}_{n;a},\mathcal{Q}_{n;b}\}=2\epsilon_{ab}
    \P_n(\mathcal{H}_n,
    {\vec{\mathcal{K}}})\mathcal{P}_{n;1}
    \,,
\end{equation}
\begin{equation}\label{Pn1QS}
     [\mathcal{P}_{n;1},\mathcal{S}_{n;a}]=i\mathcal{H}_n\P_{n-1}^-
     (\mathcal{H}_n,
    \vec{\kappa},\vec{\kappa}')\mathcal{Q}_{n;a}
    \,,\qquad
    [\mathcal{P}_{n;1},\mathcal{Q}_{n;a}]=-i\P_{n-1}^-(\mathcal{H}_n,
    \vec{\kappa},\vec{\kappa}')\mathcal{S}_{n;a}
    \,,
\end{equation}
\begin{equation}\label{Pn2QS}
    [\mathcal{P}_{n;2},\mathcal{S}_{n;a}]=i\mathcal{H}_n
     \P_n^+(\mathcal{H}_n,
    \vec{\kappa},\vec{\kappa}')\mathcal{Q}_{n;a}
    \,,\qquad
   [\mathcal{P}_{n;2},\mathcal{Q}_{n;a}]=-i\P_n^+(\mathcal{H}_n,
    \vec{\kappa},\vec{\kappa}')\mathcal{S}_{n;a}
    \,,
\end{equation}
where $\P_n(\mathcal{H}_n,\vec{\kappa})=
\prod_{j=1}^n(\mathcal{H}_n+ \kappa_j^2\cdot\mathds{1})$,
$\P_n^+(\mathcal{H}_n, \vec{\kappa},\vec{\kappa}')=
\P_n(\mathcal{H}_n,\vec{\kappa})+
\P_n(\mathcal{H}_n,\vec{\kappa}')$, $\P_{n-1}^-(\mathcal{H}_n,
\vec{\kappa},\vec{\kappa}')= \P_n(\mathcal{H}_n,\vec{\kappa})-
\P_n(\mathcal{H}_n,\vec{\kappa}'),$
$\P_n(\mathcal{H}_n,{\vec{\mathcal{K}}})=
\prod_{j=1}^n(\mathcal{H}_n+ \mathcal{K}_j^2),$
$\mathcal{K}_j=\text{diag}\,(\kappa_j',\kappa_j)$.

Operator $\P_{n-1}^-(\mathcal{H}_n, \vec{\kappa},\vec{\kappa}')$ is
a polynomial of order $n-1$ in the extended Hamiltonian
$\mathcal{H}_n$ that vanishes for $\vec{\kappa}=\vec{\kappa}'$. Then
Eq.  (\ref{Pn1QS}) signals that the supersymmetric structure of the
$n>1$ reflectionless system $\mathcal{H}_n$ with exact
isospectrality simplifies as in the case $n=1$: the integral
$\mathcal{P}_{n,1}$ turns  into bosonic central charge of the
nonlinear superalgebra. Moreover, from the form of polynomial in
$\mathcal{H}_n$ coefficients in superalgebra, one can expect that
the supersymmetric structure should undergo some radical changes
even in the case when not all the pairs of the scaling parameters
coincide but only part of them. For instance, if
$\kappa'_{j'}=\kappa_j$ for some indexes $j'$ and $j$, which may
coincide, $j'=j$, or may be different, $j'\neq j$, the same factor
 $(\mathcal{H}_n+\kappa_j^2\cdot \mathds{1})$, or its square,
appears in all the structure coefficients of the superalgebra. By
analogy with the $n=1$ case this indicates that some fermionic
supercharges may be substituted for supercharges of a lower
differential order. To understand what changes the supersummetric
structure undergoes in the cases of a partially broken or exact
isospectrality, we investigate in detail the extended system
(\ref{Hnsusy}) for the case of $n=2$ in the next Section.

\section{Supersymmetry of the $n=2$ extended
system}\label{generaln=2}

Explicit form of the supersymmetric structure for extended $n=2$
system with completely broken  isospectrality follows as a
particular case from a generic consideration of the previous
Section. Before analyzing  the partially broken and exact
isospectrality cases, we first discuss some properties of the
$n=2$ reflectionless system of the most general form. It is a
particular case of such a system, described by the two-soliton
P\"oschl-Teller Hamiltonian, that appears in $\varphi^4$ field
theoretical model with a double well potential, where it controls
the stability of the kink and anti-kink solutions.

\subsection{Generic reflectionless system with two bound states}\label{n2generic}

Explicit form of the Hamiltonian of an $n=2$  reflectionless system
of a general form is
\begin{eqnarray}\label{H2exp}
    H_{2}\left(\vec\kappa,\vec\tau\right)&=&-\frac{d^2}{dx^2} +
    V_2(x;\vec{\kappa},\vec{\tau}),\\
    V_2(x;\vec{\kappa},\vec{\tau})&=&
    -2(\kappa^2_2-\kappa_1^2)^{-1}\left(\kappa^2_2
    \text{csch}^2\kappa_2(x+\tau_2)
    +\kappa^2_1\text{sech}^2
    \kappa_1(x+\tau_1)\right)w^2(x;\vec{\kappa},\vec{\tau})\,,\label{V2}
\end{eqnarray}
where
\begin{equation}\label{wx}
    w(x;\vec{\kappa},\vec{\tau})=
    (\kappa_1^2-\kappa_2^2)\left(\kappa_2\coth{\kappa_2(x+\tau_2)}
    -\kappa_1\tanh{\kappa_1(x+\tau_1)}\right)^{-1}\,.
\end{equation}
In the limit $\tau_2\rightarrow\pm \infty$,  the two-soliton system
(\ref{H2exp}) transforms into that of the one-soliton case,
\begin{equation}\label{V2V1t2}
    V_2\rightarrow
    -2\kappa_1^2\text{sech}^2\kappa_1(x+\tau_1\mp \xi_1)\,,
\end{equation}
where a shift parameter is defined by a relation
$\sinh\kappa_1\xi_1=\kappa_1/\sqrt{\kappa_2^2-\kappa_1^2}$.
 In
another limit, $\tau_1\rightarrow\pm \infty$, the two-soliton
potential transforms into the one-soliton potential given by an
expression of the form (\ref{V2V1t2}) but with the index $1$ in the
parameters changed for $2$;  the shift parameter $\xi_2$ is given
then by a relation
$\sinh\kappa_2\xi_2=\kappa_2/\sqrt{\kappa_2^2-\kappa_1^2}$. The
indicated limits correspond to a picture of a two-soliton scattering
described by the KdV equation, where the $n=1$ solitons of
amplitudes $2\kappa_1^2$ and $2\kappa_2^2$ in such a process suffer
asymptotically only temporal shifts \cite{KDV2}.

Non-degenerate bound states  of the system (\ref{H2exp}),
${\psi}_{2}^{-\kappa^2_j}=\A_2\tilde{\psi}_j$, $j=1,2$, of energies
$E=-\kappa_1^2$ and $E=-\kappa_2^2$
 are obtained
from the partners, $\tilde{\psi_1}=\sinh{\kappa_1(x+\tau_1)}$ and
$\tilde{\psi}_2=\cosh \kappa_2(x+\tau_2)$,  of non-physical
eigenstates $\psi_1=\cosh{\kappa_1(x+\tau_1)}$ and
$\psi_2=\sinh{\kappa_2(x+\tau_2)}$ of $H_0$  by applying to them the
second order composite operator
\begin{equation}\label{A2comp}
\A_2(\vec{\kappa},\vec{\tau})=
A_{2}(\vec\kappa,\vec\tau)A_1(\kappa_1,\tau)\,,
\end{equation}
\begin{equation}\label{psi12d}
    {\psi}_{2}^{-\kappa^2_1}=
    \kappa_1 \sech\kappa_1(x+\tau_1)\,
    w(x;\vec{\kappa},\vec{\tau})\,,\qquad
    {\psi}_{2}^{-\kappa^2_2}=
    -\kappa_2\text{csch}\,\kappa_2(x+\tau_2)\,
    w(x;\vec{\kappa},\vec{\tau})\,.
\end{equation}
Here
\begin{equation}
     A_{2}(\vec\kappa,\vec\tau)=
    (A_{1}\psi_2)\frac{d}{dx}\frac{1}{(A_{1}
    \psi_2)}=-A_1^\dagger (\kappa_1,\tau_1) +
    w(x;\vec{\kappa},\vec{\tau})\,,\label{A2exp}
\end{equation}
and $A_1$ is defined by Eq. (\ref{A1def}). Function (\ref{wx})
satisfies the identities
\begin{eqnarray}\label{wV2}
    &{d w}/{dx}=\frac{1}{2}V_2\,,\\
    &w^2+2\kappa_1 w\tanh\kappa_1(x+\tau_1)=\frac{1}{2}V_2
    +\kappa_2^2-\kappa_1^2\,,&\label{wV2t}\\
    &w^2+2\kappa_2 w\coth\kappa_2(x+\tau_2)=\frac{1}{2}V_2
    +\kappa_1^2-\kappa_2^2\,,&\label{wV2c}
\end{eqnarray}
which will play a fundamental role in what follows.

The degenerate pairs of the states of the continuous part of the
spectrum with $E=k^2>0$ are obtained from the plane wave states of
the free particle, $\psi_2^{\pm k}=\A_2e^{\pm ikx}$,
\begin{eqnarray}\label{contE}
    \psi_2^{\pm k}
    =\left[-(k^2+\kappa^2_1)+(\pm
    ik-\kappa_1\tanh{\kappa_1(x+\tau_1)})w(x;\vec{\kappa},\vec{\tau})
    \right]e^{\pm
    ikx}\,.
\end{eqnarray}
The boundary case $k=0$ gives a non-degenerate,  zero energy edge
state $\psi_2^0$ at the bottom of the continuous spectrum.

The particular case of  reflectionless  $n=2$ P\"oschl-Teller
system,
$$
H_2(\kappa,\tau)=-\frac{d^2}{dx^2}-6\kappa^2\sech^2\kappa(x+\tau),
$$
is obtained by putting $\kappa_2=2\kappa_1=2\kappa$ and
$\tau_2=\tau_1=\tau$. In this case, the function (\ref{wx}) and the
operator (\ref{A2exp}) are reduced to $w=-3\kappa\tanh\chi$ and
$A_2=\frac{d}{dx}-2\kappa\tanh\chi$, the indicated bound states are
transformed, modulo overall multiplicative constants, into
$\psi_2^{-\kappa^2}=\sinh \chi\,\text{sech}^2\chi$ ($E=-\kappa^2$)
and $\psi_2^{-4\kappa^2}=\,\text{sech}^2\chi$ ($E=-4\kappa^2$),
while the zero energy non-degenerate state is $\psi_2^0=1-3\tanh^2
\chi$, where we use the notation $\chi=\kappa(x+\tau)$.

\subsection{Generalized Crum-Darboux transformations
scheme}\label{generCD}

We have constructed a generic $n=2$ reflectionless Hamiltonian
(\ref{H2exp}) by employing a sequence of two Darboux
transformations described in Section \ref{SecCD},  namely,  by
using first the non-physical free particle state
$\psi_1=\cosh\kappa_1(x+\tau_1)$, and then the state
$\psi_2=\sinh\kappa_2(x+\tau_2)$. The same final result also can
be achieved with the interchanged order of the indicated states.
This corresponds to the alternative factorization of the second
order operator (\ref{A2comp}),
\begin{equation}\label{A2BB}
    \A_2=B_2B_1\,,
\end{equation}
which intertwines $H_2$ with the free particle Hamiltonian,
$\A_2H_0=H_2\A_2$, $\A_2^\dagger H_2=H_0\A_2^\dagger$, see Fig.
\ref{fig2}b, c. The first order operators $B_1$ and $B_2$ are
obtained from $A_1$ and $A_2$ via the substitution
\begin{equation}\label{changekt}
    \kappa_1\leftrightarrow\kappa_2\,,\quad
    \tau_1\rightarrow\tau_2+i\frac{\pi}{2\kappa_2}=\tilde{\tau}_2\,,\quad
    \tau_2\rightarrow\tau_1+i\frac{\pi}{2\kappa_1}=\tilde{\tau}_1\,.
\end{equation}
This substitution leaves  invariant the Hamiltonian (\ref{H2exp}),
the second order intertwining operator $\A_2$, and the function
(\ref{wx}). It also leaves invariant the states (\ref{contE}) of the
continuous spectrum, including the non-degenerate edge state of zero
energy, but interchanges the bound states (\ref{psi12d}),
$\psi_2^{-\kappa_1^2}\rightarrow i\psi_2^{-\kappa_2^2}$,
$\psi_2^{-\kappa_2^2}\rightarrow i\psi_2^{-\kappa_1^2}$.
Transformation (\ref{changekt})  changes, however, the first order
intertwining operators $A_1$ and $A_2$, which are regular on $\R^1$,
for the singular first order operators
\begin{equation}\label{B12}
    B_1=B_1(\kappa_2,\tau_2)=\frac{d}{dx} -\kappa_2\coth
    \kappa_2(x+\tau_2)\,,\quad
    B_2=B_2(\vec{\kappa},\vec{\tau})=-B_1^\dagger(\kappa_2,\tau_2)+
    w(x;\vec{\kappa},\vec{\tau})\,.
\end{equation}
 In terms of the first order operators (\ref{B12}) we
have $H_0=B_1^\dagger B_1-\kappa_2^2$,
$\tilde{H}_1=H_1(\kappa_2,\tilde{\tau}_2)=
B_1B_1^\dagger-\kappa_2^2=B_2^\dagger B_2-\kappa_1^2$, and $H_2=
B_2B_2^\dagger-\kappa_1^2$. This means that with the alternative
factorization (\ref{A2BB}), the operator $\A_2$ intertwines
Hamiltonian (\ref{H2exp}) with $H_0$   via the $n=1$ system
described by a \emph{singular} Hamiltonian
\begin{equation}\label{H1virt}
    \tilde{H}_1=H_1(\kappa_2,\tilde{\tau}_2)
    =-\frac{d^2}{dx^2}+
    \frac{2\kappa_2^2}{\sinh^2\kappa_2(x+\tau_2)}\,.
\end{equation}
In what follows,  singular Hamiltonian (\ref{H1virt})  will appear
only as a virtual, or intermediate system, and the described
generalization of the Crum-Darboux scheme will allow us to
identify nontrivial intertwining operators for $n=2$ extended
system with partially broken and exact isospectrality. The picture
with the alternative factorizations generalizes for the case
$n>2$. In this context it is worth to note that the change of the
order of the free particle non-physical states (\ref{psij}) in the
construction of a reflectionless system $H_n$, in comparison with
that described in Section \ref{SecCD}, corresponds to a certain
permutation of the columns of the Wronskian (\ref{Wron-n}). This
produces no effect for potential in equation (\ref{lnW}).

To conclude the discussion of the generalized Crum-Darboux
transformations scheme, we present here the relations which are
helpful for computation of the
corresponding superalgebraic
structures,
\begin{eqnarray}\label{X1B1}
    &\breve{X}_{1}\left(\kappa,\tilde{\tau},
    \tilde{\tau}'\right)
     B'_1=B_1
     A_{\mathcal{C}}(\kappa,\tau-\tau')\,,\qquad
    B_1^\dagger\breve{X}_{1}
    \left(\kappa,\tilde{\tau},
    \tilde{\tau}'\right)=
    A_{\mathcal{C}}(\kappa,\tau-\tau')
    B'^\dagger_1\,,&\\
    \label{X1A1B1}
    &\breve{X}_{1}\left(\kappa,\tilde{\tau},
    \tau'\right)
     A'_1=B_1
     A_{\mathcal{C}}(\kappa,\tilde{\tau}-\tau')\,,\qquad
    B_1^\dagger\breve{X}_{1}
    \left(\kappa,\tilde{\tau},
    \tau'\right)
    =
    A_{\mathcal{C}}(\kappa,\tilde{\tau}-\tau')
      A'^\dagger_1\,,&\\
    \label{dBA}
    &A_{\mathcal{C}}(\kappa,\tau-\tilde{\tau}')
    B'^\dagger_1=
    A_1^\dagger\breve{X}_1(\kappa,\tau,\tilde{\tau}')\,,&
\end{eqnarray}
where $A_1=A_1(\kappa,\tau)$, $A'_1=A_1(\kappa,\tau')$,
$B_1=B_1(\kappa,\tau)$, $B_1'=B_1(\kappa,\tau')$,
$\tilde{\tau}=\tau+i\frac{\pi}{2\kappa}$, and
$\tilde{\tau}'=\tau'+i\frac{\pi}{2\kappa}$. These identities can be
obtained from (\ref{XbrA}) via the substitution (\ref{changekt}).

\subsection{Generic case of partial isospectrality breaking}

Now we are in position to discuss the supersymmetric structure of
the extended $n=2$ systems with partially broken and exact
isospectralities.  We first consider three cases of partial
isospectrality breaking, in which one discrete energy level
$-\kappa_j^2 $ of the subsystem $H_2$ coincides with any of the
two discrete energy levels $-\kappa_{j'}^2 $ of the partner
Hamiltonian $H_2'$, but the corresponding translation parameters
are different, $\tau_j\neq \tau_{j'}'$. All these  cases are
described by a similar supersymmetric structure. Then, in the next
subsection, we analyze the superalgebraic structure of the same
three cases but with coinciding associated translation parameters,
$\tau_j= \tau_{j'}'$.

\begin{figure}[h!]\begin{center}
\includegraphics[scale=1.2]{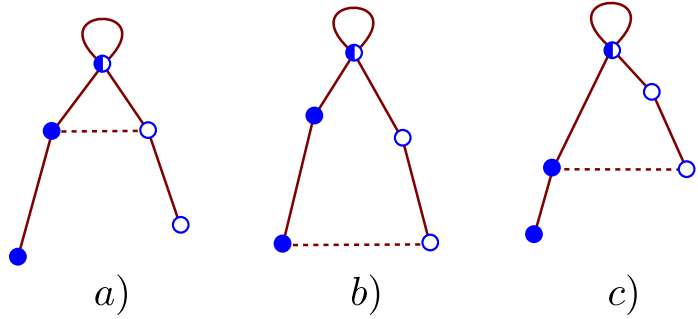}
\caption{The $n=2$ pairs with partially broken isospectrality.
}\label{fig3}
\end{center}
\end{figure}

We start with the case of partial isospectrality  breaking
characterized by the conditions
\begin{equation}\label{break1}
    \hskip-3cm \bullet\qquad\qquad
    \kappa_1=\kappa_1',\quad
    \tau_1\neq \tau_1',\qquad \kappa_2\neq
    \kappa_2',\quad \text{no restrictions on}\,\,
    \tau_2,\,  \tau_2'\,,
\end{equation}
see Fig. \ref{fig3}a.

 The subsystems
$H_2=H_2(\kappa_1,\tau_1,\kappa_2,\tau_2)$ and
$H_2'=H_2(\kappa_1,\tau_1',\kappa_2',\tau_2')$ of the extended
matrix Hamiltonian $\mathcal{H}_2$  are related  by irreducible
intertwining operators of orders $4$ and $3$, $Y_4H_2'=H_2Y_4$,
 $Y_4^\dagger H_2=H_2'Y_4^\dagger$, $\breve{X}_3^AH_2'=H_2\breve{X}_3^A$,
$\breve{X}_3^A{}^\dagger H_2=H_2'\breve{X}_3^A{}^\dagger$.  $Y_4$ is
given, in correspondence with the generic form (\ref{Yndef}), by
$Y_4=\A_2\A_2'^\dagger$, while
\begin{equation}\label{X3red}
    \breve{X}_3^A=A_2(\kappa_1,\kappa_2;\tau_1,\tau_2)
    \breve{X}_1(\kappa_1;\tau_1,\tau_1')A_2^\dagger
    (\kappa_1,\kappa_2';\tau_1',\tau_2')=
    A_2\breve{X}_1{A_2'}^\dagger\,
\end{equation}
appears instead of the fifth order intertwining operator
$X_5=\A_2\frac{d}{dx}\A_2'^\dagger$ because of the reduction
\begin{equation}\label{X5redX3A}
    X_5=(H_2+\kappa_1^2)
    \breve{X}_3^A-\mathcal{C}(\kappa_1,\tau_1-\tau_1')
    Y_4\,.
\end{equation}
As follows from (\ref{X3red}), the reduction (\ref{X5redX3A}) is
related to the opening of a `tunneling channel' via the virtual
isospectral pair of $n=1$ systems $H_1(\kappa_1,\tau_1)$ and
$H_1(\kappa_1,\tau_1')$.

Taking the products of the described intertwining operators and Lax
integrals of order $5$, $Z_5=\A_2 \frac{d}{dx}\A_2^\dagger$,
$Z'_5=\A'_2 \frac{d}{dx}\A_2'^\dagger$, $[Z_5,H_2]=0$,
$[Z_5',H_2']=0$,   presented in Appendix,  we find the
superalgebraic structure of the system $\mathcal{H}_2$ with
partially broken isospectrality (\ref{break1}). It is displayed
below in the form that unifies (\ref{break1}) with two other similar
cases.

\vskip0.1cm

A partial isospectrality breaking  with coinciding ground state
energy levels,
\begin{equation}\label{break2}
    \hskip-3cm \bullet\qquad\qquad     \kappa_2=\kappa_2',\quad
    \tau_2\neq \tau_2',\qquad \kappa_1\neq
    \kappa_1',\quad \text{no restrictions on}\,\,
    \tau_1,\,  \tau_1'\,,
\end{equation}
is similar to the previous case, see Fig. \ref{fig3}b. Intertwining
operator $Y_4$ and integrals $Z_5$ and $Z_5'$ are given by generic
formulae with restriction (\ref{break2}). The third order
irreducible intertwining operator can be obtained from (\ref{X3red})
via the substitution (\ref{changekt}),
\begin{equation}\label{brXB}
    \breve{X}_3^B=B_2\breve{X}_1(\kappa_2;\tilde{\tau}_2,\tilde{\tau}_2')
    {B_2'}^\dagger\,,
\end{equation}
where $B_2$ and $B_2'$ are given by Eq. (\ref{B12}) with
$\kappa_2'=\kappa_2$, while
\begin{eqnarray}
    \breve{X}_1(\kappa_2;\tilde{\tau}_2,\tilde{\tau}_2')&=&
    \frac{d}{dx}-\kappa_2\coth\kappa_2(x+\tau_2)+
    \kappa_2\coth\kappa_2(x+\tau_2')+\mathcal{C}(\kappa_2,\tau_2-\tau_2')
    \nonumber\\
    &=&B_1(\kappa_2,\tau_2)-B_1^\dagger(\kappa_2,\tau_2')+
    A_{\mathcal{C}}^\dagger(\kappa_2,\tau_2-\tau_2')
    \,.\label{brXB1}
\end{eqnarray}
Though all the three first order operators  that appear in
factorization of $\breve{X}_3$ in (\ref{brXB}) are singular, the
third order intertwining operator itself is regular on $\R^1$. This
follows just from the reduction relation for the fifth order
intertwining operator for the case (\ref{break2}) under
consideration,
\begin{equation}\label{X5redX3B}
    X_5=(H_2+\kappa_2^2)\breve{X}_3^{B}-
    \mathcal{C}(\kappa_2,\tau_2-\tau_2')Y_4\,.
\end{equation}
The third order intertwining operator (\ref{brXB}) realizes the
intertwining between $H_2$ and $H'_2$ by means of a `tunneling
channel' via a pair of singular $n=1$ Hamiltonians
$H_1(\kappa_2,\tilde{\tau}_2)$ and $H_1(\kappa_2,\tilde{\tau}'_2)$
of the form  (\ref{H1virt})~\footnote{By shifting the argument
$x\rightarrow x+i\delta$, where $\delta$ is a real constant, one can
translate all the consideration for the case of
$\mathcal{PT}$-symmetric quantum systems \cite{PTrev}  with
$\tilde{H}_1$ and $\tilde{H}'_1$  to be regular isospectral
Hamiltonians, see \cite{PT1}.}.

The supersymmetric structure for partial isospectrality breaking
\begin{equation}\label{break3}
    \hskip-4cm \bullet\qquad\qquad     \kappa_1=\kappa_2',\qquad \kappa_2\neq
    \kappa_1',\quad \text{no restrictions on}\,\,
    \tau_{1,2},\,  \tau_{1,2}'\,,
\end{equation}
see Fig. \ref{fig3}c, is generated in a similar way.  Here, the
third order irreducible intertwining operator is
\begin{equation}\label{brXAB}
    \breve{X}_3^{AB}=A_2\breve{X}_1(\kappa_1;\tau_1,\tilde{\tau}_2')
    {B_2'}^\dagger\,,
\end{equation}
where $B_2'=B_2(\kappa_1',\kappa_2,\tau_1',\tau_2')$ is given by Eq.
(\ref{B12}), and $\tilde{\tau}_2'=\tau_2'+i\frac{\pi}{2\kappa_1}$.
In this case, we have a reduction relation
\begin{equation}\label{X5redX3AB}
    X_5=(H_2+\kappa_1^2)\breve{X}_3^{AB}-
    \mathcal{C}(\kappa_1,\tau_1-\tilde{\tau}_2')Y_4\,.
\end{equation}
Unlike the two previous cases,
$\mathcal{C}(\kappa_1,\tau_1-\tilde{\tau}_2')=\kappa_1\tanh\kappa_1(\tau_1-\tau_2')$
is regular for any values of $\tau_1$ and $\tau_2'$  associated with
coinciding scaling parameters~\footnote{The operator (\ref{brXAB})
intertwines $H'_2$ and $H_2$ via the virtual $n=1$ systems
$\tilde{H}'_1$ and $H_1$ of different, singular and regular, nature.
After the imaginary shift mentioned in the previous footnote, the
latter pair will transform into regular $n=1$ reflectionless
P\"oschl-Teller $\mathcal{PT}$-symmetric Hamiltonians.}.

The superalgebra for the described three cases of partial
isospectrality breaking can be  presented in a unified  form
\begin{eqnarray}\label{n2SSp}
    &\{\breve{S}_{a},\breve{S}_{b}\}=
    2\delta_{ab}h_{d}h_{d'}h_{\mathcal{C}_l}\,,
    \qquad
    \{Q_{a},Q_{b}\}=
    2\delta_{ab}h^2_{i}
    h_{d}h_{d'}\,,&\\
\label{n2SQp}
    &\{\breve{S}_{a},Q_{b}\}=
    2\delta_{ab}\mathcal{C}_{l}h_{i}
    h_{d}h_{d'}+
    2\epsilon_{ab}h_{d',d}
    \mathcal{P}_{1}\,,&\\
\label{n2SP1p}
    &[\mathcal{P}_{1},\breve{S}_{a}]=
    i(\kappa^2_d-\kappa'^2_d)(h_{\mathcal{C}_l}Q_{a}-
    \mathcal{C}_{l}h_{i}\breve{S}_{a}
    )\,,
    \quad
    [\mathcal{P}_{1},Q_{a}]=
    i(\kappa^2_d-\kappa'^2_d)h_{i}(
    \mathcal{C}_{l}Q_{a}
    -h_{i} \breve{S}_{a}
    ),\qquad
    &\\
\label{n2SP2p}
    &[\mathcal{P}_{2},\breve{S}_{a}]=i(h_{d}+h_{d'})
    (h_{\mathcal{C}_l}Q_{a} -
    \mathcal{C}_{l}h_{i}\breve{S}_{a})\,,
    \quad
    [\mathcal{P}_{2},Q_{a}]=i(h_{d}+h_{d'})
    h_{i} (\mathcal{C}_{l}
    Q_{a}-h_{i}\breve{S}_{a})\,.\quad&
\end{eqnarray}
Here $h_i=\mathcal{H}_2+\kappa_i^2$, $h_d=\mathcal{H}_2+\kappa_d^2$,
$h_{d'}=\mathcal{H}_2+\kappa'^2_{d}$, $h_{d',d}=
    \mathcal{H}_2+\text{diag}(\kappa'^2_d,
    \kappa^2_d)$,
$h_{\mathcal{C}_l}=\mathcal{H}_2+\mathcal{C}_l^2$, $l=1,2,3$,
$\kappa_i$ denotes the coinciding scaling parameter of the pair,
$\kappa_d$ and $\kappa_d'$ correspond to other, not coinciding
scaling parameters of the subsystems $H_2$ and $H_2'$,
respectively, while
$\mathcal{C}_1=\mathcal{C}(\kappa_1,\tau_1-\tau_1')$ for
(\ref{break1}),
$\mathcal{C}_2=\mathcal{C}(\kappa_2,\tau_2-\tau_2')$ for
(\ref{break2}), and
$\mathcal{C}_3=\mathcal{C}(\kappa_1,\tau_1-\tilde{\tau}_2')$ for
the case (\ref{break3}). Notation $\breve{S}_{a}$ reflects the
reduction $S_{2;a}=(\mathcal{H}_2+\kappa_i^2)\breve{S}_{a}
-\mathcal{C}_lQ_{a}$ of the supercharges constructed in terms of
$X_5$ and $X_5^\dagger$, and to simplify notations, we do not
supply the supercharges with index $l$, and omitted  the index
$n=2$ in all the integrals.

The fact of a partial isospectrality breaking is reflected here in
the superalgebraic structure. On the one hand, relations
(\ref{n2SSp}), (\ref{n2SQp}) and  (\ref{n2SP2p}) are similar to
superalgebraic structure (\ref{SSbreve}), (\ref{QSbreve}) and
(\ref{PSQbreve})  of the $n=1$ isospectral case. At the same time,
the commutators  in (\ref{n2SP1p}), being  of the nature of those in
(\ref{P11QS}) for the $n=1$ non-isospectral family of the systems,
show that a `non-centrality' character of the Lax matrix integral
$\mathcal{P}_{2;1}$ is measured by the scale
 of isospectrality breaking, $\kappa_d^2-\kappa_d'^2$.

\subsection{Partial isospectrality breaking with coinciding
associated translation parameters}

\begin{figure}[h!]\begin{center}
\includegraphics[scale=1.2]{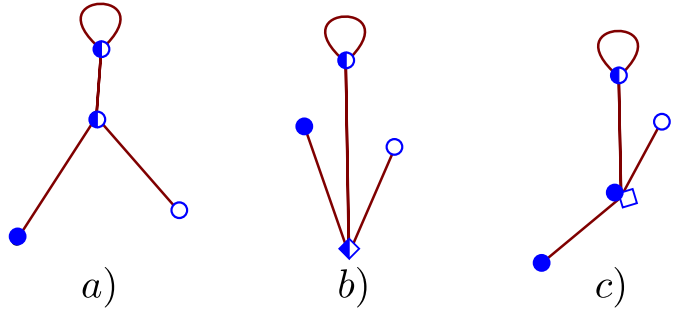}
\caption{The pairs with partially broken isospectrality, in which
the translation parameters associated with the equal scaling
parameters do coincide. In the case a), a common virtual system
corresponds to a regular $n=1$ reflectionless P\"oschl-Teller
system. In the case b) such a common virtual system is singular. In
the case c), the partners can be intertwined via a pair of $n=1$
virtual systems, one of which is singular.}\label{fig4}
\end{center}
\end{figure}

Let us discuss  now  the supersymmetry of the systems with partial
isospectrality breaking, in which one discrete energy level,
$\kappa_j=\kappa_{j'}'$,  and the associated translation parameters,
$\tau_j=\tau_{j'}'$, coincide, see Fig. \ref{fig4}a, b, c.
The two cases corresponding to either $\kappa_1=\kappa_1'$  or
$\kappa_2=\kappa_2'$ are similar. For them, supersymmetry undergoes
restructuring,  and is generated by intertwining operators of the
second, $\breve{Y}_2$, and fifth, $X_5$, orders, and by the fifth
order integrals $Z_5$ and $Z_5'$. The fifth order operators, $X_5$
and $Z_5$, in this case include in their structure the third order
integral of the corresponding common virtual $n=1$ system.

For the sake of definiteness, consider the case
$\kappa_1=\kappa_1',$ $\tau_1= \tau_1',$ $\kappa_2\neq
    \kappa_2'$.
We have $X_5=A_2 Z_3{A_2'}^\dagger$, $Z_5=A_2 Z_3{A_2}^\dagger$,
and $Z_5'=A_2' Z_3{A_2'}^\dagger$, where
$Z_3=Z_3(\kappa_1,\tau_1)=A_1(\kappa_1,\tau_1)
\frac{d}{dx}A_1^\dagger(\kappa_1,\tau_1)$ is the third order Lax
integral for the common P\"oschl-Teller virtual system
$H_1(\kappa_1,\tau_1)$. The second order intertwining operator has
a form $\breve{Y}_2^A=A_2A_2'^\dagger$, with
$A_2=A_2(\kappa_1,\kappa_2,\tau_1,\tau_2)$ and
$A_2'=(\kappa_1,\kappa_2',\tau_1,\tau_2')$, and the fourth order
intertwining  operator ${Y}_4=\A_2{\A_2'}^\dagger$ of a generic
case reduces as
\begin{equation}\label{Y4Y2red}
    Y_4=(H_2+\kappa_1^2)\breve{Y}_2^A\,.
\end{equation}
The second order operator $\breve{Y}_2^A$ can be obtained also from
the third order operator (\ref{X3red}) of the  case (\ref{break1})
considered above. Indeed, multiplying (\ref{X3red}) by
$-\mathcal{C}^{-1}(\kappa_1,\tau_1-\tau_1')$, and taking a limit
$\tau_1'\rightarrow \tau_1$, we get $\breve{Y}_2^A$. So, the change
of supersymmetric structure is related here to a singular nature of
$\mathcal{C}(\kappa_1,\tau_1-\tau_1')$ in the limit
$\tau_1'\rightarrow \tau_1$.  Another case, with
$\kappa_2=\kappa_2',$ $\tau_2= \tau_2',$ $\kappa_1\neq    \kappa_1'$
is treated in a similar way, and superalgebraic structure for these
two cases can be presented in a unified form:
\begin{equation}\label{PaBrt1}
    \{S_{a},S_{b}\}=2\delta_{ab}\mathcal{H}_2
    h^2_{i}h_{d}h_{d'},\qquad
    \{\breve{Q}_{a},\breve{Q}_{b}\}=
    2\delta_{ab}h_{d}
    h_{d'},
\end{equation}
\begin{equation}\label{PaBRt3}
    \{S_{a},\breve{Q}_{b}\}=
    2\epsilon_{ab}h_{d',d}\mathcal{P}_{1}\,,
\end{equation}
\begin{equation}\label{PaBRt4}
    [\mathcal{P}_{1},S_{a}]
    =i(\kappa^2_d-\kappa'^2_d)
    \mathcal{H}_2h^2_{i}\breve{Q}_{a}\,,
    \quad
     [\mathcal{P}_{1},\breve{Q}_{a}]
    =-i(\kappa^2_d-\kappa'^2_d)S_{a}\,,
\end{equation}
\begin{equation}\label{PaBRt5}
    [\mathcal{P}_{2},S_{a}]=
    i
    \mathcal{H}_2 h^2_{i}
    (h_{d}+h_{d'})\breve{Q}_{a}\,,
    \quad
    [\mathcal{P}_{2},\breve{Q}_{a}]
    =-i(h_{d}+h_{d'})S_{a}\,.
\end{equation}
Notation $\breve{Q}_{a}$ reflects here the reduction
$Q_{2;a}=(\mathcal{H}_2+\kappa_i^2)\breve{Q}_{a}$, and, again, we
omitted the index $n=2$ in the specification of nontrivial
integrals.

The case $\kappa_1=\kappa_2',$
    $\tau_1= \tau_2',$ $\kappa_1\neq
    \kappa_2'$
is different from the two previous ones because the corresponding
parameter-dependent function
$\mathcal{C}(\kappa_1,\tau_1-\tilde{\tau}_2')=
\kappa_1\tanh\kappa_1(\tau_1-\tau_2')$ is non-singular for any
values of $\tau_1-\tau_2'$, and, moreover, turns into zero at
$\tau_1= \tau_2'$. Here,  the intertwining operators are $Y_4$, and
$\breve{X}_3^{AB}$ given by Eq. (\ref{brXAB}) with $\tau_1=\tau_2'$.
A non-singular nature of the latter is seen from (\ref{X5redX3AB}).
The superalgebra for this case is obtained directly from
(\ref{n2SSp})--(\ref{n2SP2p}) just by putting there
$\mathcal{C}_3=0$. Though here the irreducible intertwining
generators are different in comparison with the previous two cases,
the resulting superalgebra (\ref{n2SSp})--(\ref{n2SP2p}) with
$\mathcal{C}_3=0$ has a similar form to
(\ref{PaBrt1})--(\ref{PaBRt5}). Notice also a remarkable similarity
of (\ref{PaBrt1})--(\ref{PaBRt5}) with the superalgebra
(\ref{S1S1})--(\ref{P11QS}) of the $n=1$ non-isospectral case.

We see that in all the three cases of partial breaking of
isospectrality with corresponding coinciding translation
parameters (associated with coinciding discrete energy levels),
the superalgebraic structure does not depend on the two remaining
translation parameters associated with the second, different
discrete energy levels.

In all the cases of partial isospectrality breaking described in
this and previous subsections, the total order of the two basic
intertwining operators is the same, $3+4=2+5=7$, being less in $2$
in comparison with the complete isospectrality breaking case.

\subsection{Exact isospectrality with a common virtual $n=1$
system}\label{n2exact}

\begin{figure}[h!]\begin{center}
\includegraphics[scale=1.2]{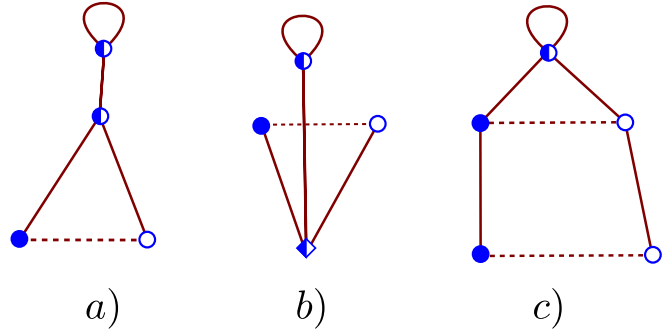}
\caption{The $n=2$ isospectral pairs with a common regular, (a), or
singular, (b), virtual system. A general case of the $n=2$
isospectral pair with $\tau_j\neq \tau'_j$, $j=1,2$, is illustrated
by c).}\label{fig5}
\end{center}
\end{figure}

The supersymmetric structure  of the systems with exact
isospectrality, $\kappa_1=\kappa_1'$ and $\kappa_2=\kappa_2'$,
depends on whether the corresponding  translation parameters are
different, $\tau_j\neq\tau_j'$, $j=1,2$, or they coincide in one of
the pairs~\footnote{The case when both pairs of translation
parameters coincide corresponds to $\mathcal{H}_2$ composed from the
two copies of the same Hamiltonian $H_2$ . Such a system
$\mathcal{H}_2$ is described by a trivial supersymmetric structure
to be similar to that discussed for $n=1$ case in Section
\ref{1selfiso}, with integral $Z_3$ changed for $Z_5$.}. The
analysis of the second case, see Fig. \ref{fig5}a, b, is more
simple, and we first consider it supposing, for the sake of
definiteness, that $\tau_1=\tau_1'$, $\tau_2\neq \tau_2'$.
The intertwining operators for such an isospectral system with a
common regular virtual $n=1$ system $H_1(\kappa_1,\tau_1)$ are
\begin{equation}\label{Y2X3ex}
    \breve{Y}_2^A=A_2A'^\dagger_2\,,\qquad
    \breve{X}_3^B=B_2
    \breve{X}_1(\kappa_2,\tilde{\tau}_2,\tilde{\tau}_2')B'^\dagger_2\,,
\end{equation}
where in $A'_2$ and $B_2'$ we assume that $\kappa'_j=\kappa_j$,
$j=1,2$, and $\tau'_1=\tau_1$, $\tau'_2\neq \tau_2$. They can be
obtained here via the reduction relations of generic intertwining
operators,
\begin{equation}\label{Y2X3exact11}
    X_5=(H_2+\kappa_2^2)\breve{X}_3^B-
    \mathcal{C}(\kappa_2,\tau_2-\tau_2')Y_4\,,\qquad
     Y_4=(H_2+\kappa_1^2)\breve{Y}_2^A\,.
\end{equation}
The intertwining generators $\breve{Y}_2^B$ and $\breve{X}_3^A$,
and the corresponding reduction relations for the exact
isospectrality case $\kappa_j=\kappa'_j$, $j=1,2$,
$\tau'_2=\tau_2$, $\tau'_1\neq \tau_1$ are obtained from
(\ref{Y2X3ex}) and (\ref{Y2X3exact11}) by changing
$\kappa_1\leftrightarrow \kappa_2$, $\tau_1\leftrightarrow
\tau_2$, $\tau'_1\leftrightarrow \tau'_2$, and $A_2\leftrightarrow
B_2$.

The nontrivial relations of superalgebraic structure for the
isospectral case with $\tau_1=\tau'_1$, $\tau_2\neq \tau'_2$ are
\begin{equation}\label{susyn=21101}
    \{\breve{\mathcal{S}}_{a},\breve{\mathcal{S}}_{b}\}
    =2\delta_{ab}h_{\mathcal{C}_2}h_{1}^2,\quad
    \{\breve{\mathcal{Q}}_{a},\breve{\mathcal{Q}}_{b}\}
    =2\delta_{ab}h_{2}^2,
\end{equation}
\begin{equation}\label{susyn=21102}
    \{\breve{\mathcal{S}}_{a},
    \breve{\mathcal{Q}}_{b}\}=2\delta_{ab}\mathcal{C}_2h_{1}
    h_{2}+2\epsilon_{ab}\mathcal{P}_{1},
\end{equation}
\begin{equation}\label{susyn=21103}
    [\mathcal{P}_{2},\breve{\mathcal{S}}_{a}]=
    2ih_{1}(h_{\mathcal{C}_2}h_{1}
    \breve{\mathcal{Q}}_{a}-\mathcal{C}_2h_{2}
    \breve{\mathcal{S}}_{a}),\quad
     [\mathcal{P}_{2},\breve{\mathcal{Q}}_{a}]=
    2ih_{2}(\mathcal{C}_2 h_{1}
    \breve{\mathcal{Q}}_{a}-h_{2}
    \breve{\mathcal{S}}_{a}),
\end{equation}
where $\mathcal{C}_2=\kappa_2\coth\kappa_2(\tau_2-\tau'_2)$,
$h_{i}=\mathcal{H}_2+\kappa_i^2$, $i=1,2$, and $h_{\mathcal{C}_2}=
\mathcal{H}_2+\mathcal{C}_2^2$.
 The superalgebra for the  isospectral case with
$\tau_2=\tau'_2$, $\tau_1\neq \tau'_1$ is obtained from the
displayed one by changing $\mathcal{C}_2\rightarrow \mathcal{C}_1$,
$h_{1}\leftrightarrow h_{2}$ in the right hand side expressions. The
supersymmetry (\ref{susyn=21101}), (\ref{susyn=21102}),
(\ref{susyn=21103}) has the structure similar to that for the $n=1$
isospectral case.

As it is expected, the integral $\mathcal{P}_{2;1}$ transmutes  here
into the bosonic central charge of nonlinear superalgebra. The total
order of the basic irreducible intertwining generators reduces in
two in comparison with the partially broken isospectrality case  and
equals the order $5$ of Lax integrals $Z_5$ and $Z'_5$. In
correspondence with this, the anticommutator of the second order,
$\breve{\mathcal{Q}}_{2;a}$, and the third order,
$\breve{\mathcal{S}}_{2;a}$, supercharges taken with different
values of indexes $a$ and $b$ is equal to the central charge
$\mathcal{P}_{2;1}$ up to a numerical, Hamiltonian-independent,
coefficient, see Eq. (\ref{susyn=21102}). The superalgebraic
structure also detects the difference of the non-coinciding
translation parameters.

\subsection{Generic case of $n=2$ exact
isospectrality}\label{n=2exactiso}

Consider a generic case of exact isospectrality characterized by
the relations $\kappa_1=\kappa'_1$, $\kappa_2=\kappa'_2$,
$\tau_1\neq \tau'_1$, $\tau_2\neq \tau'_2$, see Fig. \ref{fig5}c.
The second order intertwining operator $X_5$ possesses then two
distinct reductions, (\ref{X5redX3A}) and (\ref{X5redX3B}), in
which  it is necessary to put in addition, respectively,
$\kappa_2=\kappa'_2$ and $\kappa_1=\kappa'_1$. The existence of
the two third order intertwining operators means that a generic
isospectral case is described  by the basic intertwining operators
of the orders $2$ and $3$, to which the intertwining operators
$X_5$ and $Y_4$ are reduced. To see this, we note that
$\breve{X}_3^A$ and $\breve{X}_3^B$ are the third order operators
with the same coefficient $-1$ before the leading derivative term.
Then the difference of these two operators has to be an
intertwining differential operator of the second order.  This
implies that the coefficient before the leading second order
derivative term in the latter should be a constant. Taking into
account that $A_2=-A_1^\dagger+w$ and $B_2=-B_1^\dagger +w$, and
employing relations (\ref{XbrA}) and (\ref{X1B1}), we find
\begin{equation}\label{XAXBY}
    \breve{X}_3^A-\breve{X}_3^B=
    (\mathcal{C}_1-\mathcal{C}_2)G_2+(\kappa_2^2-\kappa_1^2)\hat{X}_1\,,
\end{equation}
where $\mathcal{C}_1=\kappa_1\coth\kappa_1(\tau_1-\tau'_1)$, and
$\mathcal{C}_2=\kappa_2\coth\kappa_2(\tau_2-\tau'_2)$,
\begin{equation}\label{G2}
    G_2=-\frac{d^2}{dx^2}+(w'-w)\frac{d}{dx}+\frac{dw'}{dx}
    +ww' +w\kappa_2\coth\kappa_2(x+\tau_2)+
    w'\kappa_2\coth\kappa_2(x+\tau'_2)+\kappa^2_2\,,
\end{equation}
\begin{equation}\label{hatX1}
    \hat{X}_1=\frac{d}{dx}+(w-w')+\mathcal{C}_1\,.
\end{equation}
In (\ref{G2}) and (\ref{hatX1}) $w$ corresponds to the function
(\ref{wx}), and $w'$ is the same function but with $\tau_j$ changed
for $\tau'_j$, $j=1,2$. {}From (\ref{XAXBY}) it
 follows immediately that the case $\mathcal{C}_1=\mathcal{C}_2$ is
special, and we shall consider it in the next subsection. So, till
the end of this subsection we suppose that
\begin{equation}\label{C1neqC2}
    \mathcal{C}_1\neq \mathcal{C}_2\,.
\end{equation}
We obtain then the second order intertwining operator
\begin{equation}\label{hatY2}
    \hat{Y}_2=\frac{\breve{X}_3^A-\breve{X}_3^B}{\mathcal{C}_1-\mathcal{C}_2}=
    G_2+\frac{\kappa_2^2-\kappa_1^2}{\mathcal{C}_1-\mathcal{C}_2}
    \hat{X}_1\,.
\end{equation}
The operator  (\ref{hatY2}) intertwines $H'_2$ and $H_2$,
$\hat{Y}_2H_2'=H_2\hat{Y}_2$, and satisfies the relation
$\hat{Y}_2^\dagger=\hat{Y}'_2$, where $\hat{Y}'_2$ corresponds to
$\hat{Y}_2$ with the interchanged translation parameters $\tau_j$
and $\tau'_j$, $j=1,2$. $\hat{Y}_2^\dagger$ generates the
intertwining relation in the reverse direction. Operator
$\hat{Y}_2$, and any of two third order operators, $\breve{X}_3^A$
or $\breve{X}_3^B$, play now a role of independent intertwining
generators. It is more convenient, however, to take a linear
combination
\begin{equation}\label{hatX3}
    \hat{X}_3=\frac{\mathcal{C}_2\breve{X}_3^A-
    \mathcal{C}_1\breve{X}_3^B}{\mathcal{C}_2-\mathcal{C}_1}\,,
\end{equation}
different from that in (\ref{XAXBY}), as a third order intertwining
generator to be independent from $\hat{Y}_2$. Using Eqs.
(\ref{X5redX3A}) and (\ref{X5redX3B}), we find that the generic
intertwining operators $X_5$ and $Y_4$ are reduced here as follows,
\begin{eqnarray}\label{redXYisos1}
    (\mathcal{C}_1-\mathcal{C}_2)X_5&=&\left(
    (\mathcal{C}_1-\mathcal{C}_2)H_2+\mathcal{C}_1\kappa_2^2
    -\mathcal{C}_2\kappa_1^2\right)\hat{X}_3+(\kappa_2^2-\kappa_1^2)
    \mathcal{C}_1\mathcal{C}_2\hat{Y}_2\,,\\
    (\mathcal{C}_1-\mathcal{C}_2)Y_4&=&
    (\kappa_2^2-\kappa_1^2)\hat{X}_3+
    \left(
    (\mathcal{C}_1-\mathcal{C}_2)H_2+\mathcal{C}_1\kappa_1^2
    -\mathcal{C}_2\kappa_2^2\right)\hat{Y}_2\,.
    \label{redXYisos2}
\end{eqnarray}
Proceeding from the relations (\ref{redXYisos1}),
(\ref{redXYisos2}) and the relations, presented in the Appendix,
which correspond to the products of operators $X_5$, $Y_4$ and
$Z_5$ with the imposed isospectrality relations
$\kappa_j=\kappa'_j$, $j=1,2$, one can find all the products of
the irreducible intertwining operators $\hat{Y}_2$, $\hat{X}_3$,
$\hat{Y}_2^\dagger$, $\hat{X}_3^\dagger$, and Lax operators $Z_5$
and $Z'_5$. With these, one can compute the superalgebra generated
by the second order, $\breve{\mathcal{Q}}_{2;a}$, and third order,
$\breve{\mathcal{S}}_{2;a}$, supercharges constructed in terms of
$\hat{Y}_2$ and $\hat{X}_3$ following the same rules as we used
before, and by the fifth order bosonic integrals
$\mathcal{P}_{2;a}$. There is  another, more simple way to compute
the superalgebra. Having in mind that fermionic supercharges are
matrix differentials operators of orders $2$ and $3$, the
alternative form of superalgebra is generated by taking a linear
combination of them,
$F^A_{a}=\mathcal{C}_1\breve{\mathcal{Q}}_{2;a}+\breve{\mathcal{S}}_{2;a}$
and
$F^B_{a}=\mathcal{C}_2\breve{\mathcal{Q}}_{2;a}+\breve{\mathcal{S}}_{2;a}$,
constructed from $\breve{X}_3^A$ and $\breve{X}_3^B$ in
correspondence with relations (\ref{XAXBY}) and (\ref{hatX3}),
\begin{equation}\label{FAB}
    F^{A,B}_{1}=\left(%
\begin{array}{cc}
  0 & \breve{X}^{A,B}_3 \\
  \breve{X}^{A,B\dagger}_3 & 0 \\
\end{array}%
\right),\qquad
    F^{A,B}_{2}=i\sigma_3F^{A,B}_{1}\,.
\end{equation}
Modifying further the notations, $F_a^{(1)}=F_a^A$,
$F_a^{(2)}=F_a^B$, and using the product relations of the
operators $\breve{X}^{A,B}_3$, their conjugate,
$\breve{X}^{A,B\dagger}_3$, and Lax operators $Z_5$ and $Z'_5$,
see Appendix, we present nonzero superalgebraic relations in a
compact form,
\begin{equation}\label{FF}
    \{F^{(i)}_a,F^{(j)}_b\}=
    2\delta_{ab}h_{ij}h_ih_j+2\epsilon_{ab}\epsilon^{ij}
    \Delta\mathcal{C}\,\mathcal{P}_{1},
\end{equation}
\begin{equation}\label{PF}
    [\mathcal{P}_{2},F^{(j)}_a]=
    \frac{2i}{\Delta\mathcal{C}}\left(
    (-1)^jh_1h_2h_{12}F^{(j)}_a+\epsilon^{jk}h_{jj}h_k^2
    F^{(k)}_a\right).
\end{equation}
Here $h_i=\mathcal{H}_2+\kappa_i^2$, $h_{ij}=\mathcal{H}_2+
\mathcal{C}_i\mathcal{C}_j$, $i,j=1,2$,
$\Delta\mathcal{C}=\mathcal{C}_2-\mathcal{C}_1$,  and no summation
in the indexes $i$ and $j$ is implied in the right hand sides.

Again, the integral $\mathcal{P}_{1}=\mathcal{P}_{2;1}$ transmutes
here into the bosonic central charge, and the structure coefficients
depend on both relative translation parameters via $\mathcal{C}_1$
and $\mathcal{C}_2$.

The nonzero superalgebraic relations for the third,
$\breve{\mathcal{S}}_{2;a}$, and second,
$\breve{\mathcal{Q}}_{2;a}$, order supercharges and  bosonic
integrals $\mathcal{P}_{2;2}$ can now easily be obtained from
(\ref{FF}) and (\ref{PF}) by employing the relations $
\breve{\mathcal{Q}}_{2;a}=
(F^{(2)}_a-F^{(1)}_a)/\Delta\mathcal{C}$,
$\breve{\mathcal{S}}_{2;a}=
(\mathcal{C}_2F^{(1)}_a-\mathcal{C}_1F^{(2)}_a)/\Delta\mathcal{C}$.
The superalgebra has the same structure (\ref{SSbreve}),
(\ref{QSbreve}), (\ref{PSQbreve}) as for the $n=1$ isospectral
case, but with Hamiltonian-dependent coefficients of a more
complicated form.

\subsection{Special case of isospectrality with
$\mathcal{C}_1=\mathcal{C}_2$}\label{n=2special}

Let us consider the special case of isospectrality characterized by
the relation
\begin{equation}\label{C1=C2}
    \mathcal{C}_1=\mathcal{C}_2\,.
\end{equation}
Equation (\ref{C1=C2}) means that there is a special correlation
between relative displacements $\tau_1-\tau_1'$ and $\tau_2-\tau'_2$
and scaling parameters,
$\kappa_1\coth\kappa_1(\tau-\tau'_1)=\kappa_2\coth\kappa_2(\tau_2-\tau_2')$.
In correspondence with this relation, we may take an $n=2$ system
$H_2$ defined by arbitrary parameters $\kappa_2>\kappa_1$, and
arbitrary, but finite, $\tau_1$ and $\tau_2$. Particularly, we can
choose the $n=2$  P\"oschl-Teller system defined by the relations
$\kappa_2=2\kappa_1$ and $\tau_1=\tau_2$. The  partner Hamiltonian
$H'_2$ is given then by the same scaling parameters, the finite
parameter $\tau'_2$ may be chosen in an arbitrary way with the only
restriction $\tau'_2\neq \tau_2$,  while $\tau'_1$ is fixed
uniquely, $\tau'_1=\tau_1-\frac{1}{\kappa_1}\text{arccoth}\,
(\frac{\kappa_2}{\kappa_1}\coth\kappa_2 (\tau_2-\tau'_2))$.

As a consequence of relation (\ref{XAXBY}), here a difference
$\breve{X}^A_3-\breve{X}_3^B$ reduces to the first order
intertwining operator (\ref{hatX1}), which satisfies a relation
$\hat{X}_1^\dagger(\vec{\kappa},\vec{\tau},\vec{\tau}\,')=
-\hat{X}_1(\vec{\kappa},\vec{\tau}\,',\vec{\tau})= -\hat{X}_1'$.
Moreover, we will show below that each of the third order
intertwining operators $\breve{X}_3^A$ and $\breve{X}_3^B$ is
reducible, and so, here the irreducible intertwining operators are
$\hat{X}_1$ and $Y_4$.

As the intertwining  generator $\hat{X}_1$ is the first order
differential operator, let us define a superpotential $W$ by means
of
\begin{equation}\label{XWsuper}
    \hat{X}_1=\frac{d}{dx}+W\,,\qquad
    W=w-w'+\mathcal{C}_1\,.
\end{equation}
In accordance with relations (\ref{wV2}), (\ref{wV2t}),
(\ref{wV2c}), we have $W^2+W'=V_2+\mathcal{C}_1^2$,
$W^2-W'=V_2'+\mathcal{C}_1^2$, and then
\begin{equation}\label{X1X1H}
    \hat{X}_1\hat{X}_1^\dagger=H_2+\mathcal{C}_1^2,\qquad
    \hat{X}^\dagger \hat{X}_1=H_2'+\mathcal{C}_1^2\,,
\end{equation}
and $\hat{X}_1 H'_2=H_2\hat{X}_1$, $\hat{X}_1^\dagger
H_2=H'_2\hat{X}_1^\dagger$. The first order intertwining operator
$\hat{X}_1$ has a form similar to that of the operator $\breve{X}_1$
in the $n=1$  isospectral case. The superpotential $W(x)$ plays here
a role of the gap function $\Delta$ mentioned there in the context
of its relation to the Bogoliubov-de Gennes system.

Operator $\hat{X}_1$ together with the first order operators
$\breve{X}^A_1=\breve{X}_1(\kappa_1,\tau_1,\tau_1')$,
$\breve{X}^B_1=\breve{X}_1(\kappa_2,\tau_2,\tau_2')$ satisfies in
addition the identities
\begin{eqnarray}\label{XXAA}
    &\breve{X}^A_1A_2'^\dagger=A_2^\dagger \hat{X}_1\,,
    \quad
    A_2\breve{X}^A_1=\hat{X}_1 A'_2\,,\qquad
     \breve{X}^B_1B_2'^\dagger=B_2^\dagger \hat{X}_1\,,
    \quad
    B_2\breve{X}^A_1=\hat{X}_1 B'_2\,.&
\end{eqnarray}
Let us stress that like (\ref{X1X1H}), these relations are valid
only in the  special isospectral case (\ref{C1=C2}). Employing them,
we find that the third order intertwining generators $\breve{X}^A_3$
and $\breve{X}^B_3$ are reducible,
\begin{equation}\label{XAXBred}
    \breve{X}^A_3=(H_2+\kappa_2^2)\hat{X}_1\,,\qquad
    \breve{X}^B_3=(H_2+\kappa_1^2)\hat{X}_1\,.
\end{equation}
As a consequence, the fifth order generic intertwining operator also
is reducible,
$X_5=(H_2+\kappa_1^2)(H_2+\kappa_2^2)\hat{X}_1-\mathcal{C}_1Y_4$.

Applying the product relations
(\ref{n2specialCC01})-(\ref{n2specialCC04}) collected in Appendix,
we can compute the superalgebra generated by the fermionic
supercharges $\hat{\mathcal{S}}_{2;a}$ constructed from $\hat{X}_1$
and $\hat{X}_1^\dagger$, by the supercharges $\mathcal{Q}_{2;a}$
composed from $Y_4$ and $Y_4^\dagger$, and by the bosonic integrals
$\mathcal{P}_{2;a}$ constructed from Lax operators $Z_5$ and $Z'_5$.
The nontrivial (anti) commutations relations are
\begin{eqnarray}\label{susyn2spec1}
    &\{\hat{\mathcal{S}}_{a},\hat{\mathcal{S}}_{b}\}=
    2\delta_{ab}h_{\mathcal{C}_1}\,,\qquad
    \{\mathcal{Q}_{a},\mathcal{Q}_{b}\}=
    2\delta_{ab}h_{1}^2h_{2}^2\,,&\\
    &\{\hat{\mathcal{S}}_{a},\mathcal{Q}_{b}\}=
    2\delta_{ab}\mathcal{C}_1h_{1}h_{2}+2\epsilon_{ab}
    \mathcal{P}_{1}\,,&\label{susyn2spec2}\\
    &[\mathcal{P}_{2},\hat{\mathcal{S}}_{a}]=2i
    (h_{\mathcal{C}_1}\mathcal{Q}_{a}
    -\mathcal{C}_1h_{1}h_{2}\hat{\mathcal{S}}_{a}
    )\,,\quad
    [\mathcal{P}_{2},\mathcal{Q}_{a}]=2i
    h_{1}h_{2}(\mathcal{C}_1\mathcal{Q}_{a}-
    h_{1}h_{2}\hat{\mathcal{S}}_{a}
    )\,,\qquad \label{susyn2spec3}&
\end{eqnarray}
where $h_i=\mathcal{H}_2+\kappa_i^2$, $i=1,2$,
$h_{\mathcal{C}_1}=\mathcal{H}_2+\mathcal{C}_1^2$, and we omitted
the index $n=2$ in the integrals.

Supercharges $\hat{\mathcal{S}}_{2;a}$, $a=1,2$,  generate a Lie
sub-superalgebra of $N=2$ supersymmetry. Since
$\mathcal{C}_1^2=\mathcal{C}_2^2>\kappa_2^2$, it corresponds to
the spontaneously broken phase. However, a peculiarity of the
extended system $\mathcal{H}_2$ is that it has a structure of
centrally extended $N=4$ nonlinear superasymmetry with the two
additional fourth order supercharges $\mathcal{Q}_{2;a}$, and two
bosonic integrals $\mathcal{P}_{2;a}$. Again, the integral
$\mathcal{P}_{2;1}$ plays here the role of the central charge. As
in a generic isospectral case from the previous subsection, the
sum of differential orders of the basic irreducible intertwining
operators equals $5$ and coincides with the order of Lax
operators. Again, the superalgebra (\ref{susyn2spec1}),
(\ref{susyn2spec2}), (\ref{susyn2spec3}) has a remarkable
similarity with that for the $n=1$ isospectral case.

We conclude that with a chosen subsystem $H_2$, Eq. (\ref{C1=C2})
defines a one-parametric family, in which  $\tau'_2$, , $\tau'_2\neq
\tau_2$, is a free parameter of the exactly isospectral system
$H'_2$. Such a family of the Schr\"odinger pairs  is described by
the supersymmetry with the two first order supercharges, two
supercharges of order four, and two bosonic integrals of
differential order five, one of which is a central charge. This
generalizes the $n=1$ self-isospectral case discussed in Section 4
for the case of $n=2$  isospectral, but not self-isospectral, pairs.

\section{Partially broken and exact isospectralities in $n>2$
systems}\label{nisospectrality}

The analysis of partially broken and exact isospectralities can be
generalized for $n$-soliton extended systems with $n>2$. The case
$n=2$ considered in the previous Section shows that the concrete
structure of supersymmetry, namely its irreducible generators and
coefficients in the superalgebra,  depends not only on how many
scaling parameters coincide, but also on whether they correspond to
the same or different ordinal numbers of discrete energy levels of
subsystems. It also depends on relative translation parameters
associated with the corresponding coinciding discrete energy levels,
and may change in the cases when such relative translation
parameters turn into zero, or are correlated via equalities  of the
form (\ref{C1=C2}). Correspondingly,  a concrete form of
supersymmetric structure is rather variable,  but the general
picture can be summarized as follows. The $n>2$  pair is
characterized by two irreducible basic intertwining operators, one
of which is a differential operator of odd order, while another is
of even order. Each $n$-soliton subsystem also is characterized by a
nontrivial integral to be a differential Lax operator of order
$2n+1$. The orders of irreducible intertwining operators satisfy the
following rules. As we saw, the case of complete isospectrality
breaking, when all the scaling parameters of one subsystem are
different from those of the second subsystem, the supersymmetric
pair is characterized by intertwining operators, $X_{2n+1}$ and
$Y_{2n}$, of differential orders $\vert X_{2n+1}\vert=2n+1$ and
$\vert Y_{2n}\vert=2n$. The sum of their differential orders,
$4n+1$, coincides with the order of the composite differential
operator of the form $(H_n)^n Z_n$. When any pair of the scaling
parameters of the subsystems coincides, the total order of the two
basic irreducible intertwining operators decreases in such a way
that $|XY^\dagger|=|(H_n)^{n-1}P|=4n-1$. Any new coincidence of some
new pair of scaling  parameters  decreases the total order of
$XY^\dagger$ in two. Finally, in the case of exact isospectrality,
when all the $n$ pairs of the scaling parameters coincide, we have
$|XY^\dagger|=|Z_n|=(4n+1)-2n=2n+1$.

As an example, consider a generic  case of exact isospectrality for
the pair of the reflectionless soliton systems, each having three
bound states. In this case, the composite operator $\A_3$ has $6$
different factorizations in dependence on the order of the free
particle non-physical states $\psi_j$, $j=1,2,3$, which are used to
generate a $3$-soliton system.  For instance, factorization
$\A_3=A_3^{(3)}A_2^{(2)}A_1^{(1)}$ corresponds to that described in
Section \ref{nisobrokencomp}, while
$\A_3=A_3^{(3)}A_2^{(1)}A_1^{(2)}$ corresponds to alternative
factorization like that described in Section \ref{generCD}, with
$A_1^{(2)}$ constructed in terms of the state $\psi_2$, $A_2^{(1)}$
constructed recursively in terms of $A_1^{(2)}$ and $\psi_1$, and
finally, $A_3^{(3)}$  is constructed recursively by employing
$A_1^{(2)}$, $A_2^{(1)}$ and  $\psi_3$. In other words,  the upper
index indicates here the index of a state $\psi_j$ we use to
construct the first order Darboux operator of the generation marked
by the lower index. The factorizations different from the standard
one $\A_3=A_3^{(3)}A_2^{(2)}A_1^{(1)}$ correspond to permutations of
columns in the Wronskian (\ref{Wron-n}), and in accordance with Eq.
(\ref{lnW}), do not produce any effect on the final form of the
three-soliton potential $V_3$. Employing the information on
intertwining operators of the $n=2$ case, we construct three
intertwining operators of order $5$,
$\breve{X}_5^{(1)}=A_3^{(3)}\breve{X}_3^{(12)}A'^{(3)\dagger}_3$,
$\breve{X}_5^{(2)}=A_3^{(1)}\breve{X}_3^{(23)}A'^{(1)\dagger}_3$,
and
$\breve{X}_5^{(3)}=A_3^{(2)}\breve{X}_3^{(31)}A'^{(2)\dagger}_3$,
where $\breve{X}_3^{(12)}=A_2 \breve{X}_1^{(1)}A'^{(2)\dagger}_2$
and
$\breve{X}_1^{(1)}=A_1^{(1)}-A'^{(1)\dagger}_1-A_{\mathcal{C}_1}$ is
the first order operator constructed in accordance with Eq.
(\ref{X1breve}), and
$\mathcal{C}_r=\kappa_r\coth\kappa_r(\tau_r-\tau'_r)$, $r=1,2,3$.
The generic intertwining operator  of order $7$ reduces as
\begin{equation}\label{X7}
    X_7=(H_3+\kappa_r^2)\breve{X}_5^{(r)}-\mathcal{C}_rY_6\,,\quad
    r=1,2,3\,.
\end{equation}
Taking $(\breve{X}_5^{(1)}-\breve{X}_5^{(2)})$ and
$(\breve{X}_5^{(2)}-\breve{X}_5^{(3)})$, we get two intertwining
operators of order $4$, $\breve{Y}^{(12)}$ and $\breve{Y}^{(23)}$,
in which the coefficients before leading derivative term $d^4/dx^4$
will be constants. Presenting  $\breve{Y}^{(12)}$ and
$\breve{Y}^{(23)}$ in a normal form, with leading coefficients to be
equal to $1$, and taking a difference of the resulting fourth order
differential operators, we get irreducible intertwining operator of
order $3$. Taking any one of the obtained two fourth order
operators, we identify finally a pair of the basic irreducible
intertwining operators $\hat{X}_3$ and $\hat{Y}_4$ of orders $3$ and
$4$. Three identities in (\ref{X7}) allow us then, on the one hand,
to express the generic intertwining operators $X_7$ and $Y_6$, which
are reducible here,  in terms of $\hat{X}_3$ and $\hat{Y}_4$
multiplied by certain polynomials in $H_3$. On the other hand, the
same identities (\ref{X7}) indicate that the cases with
$\mathcal{C}_1=\mathcal{C}_2$ and/or $\mathcal{C}_2=\mathcal{C}_3$
are peculiar. Coherently with the analysis of the previous Section,
one can expect that in the special case
$\mathcal{C}_1=\mathcal{C}_2=\mathcal{C}_3$ the basic irreducible
intertwining operators are of orders $1$ and $6$. The analysis of
this special case  requires a separate consideration and we do not
present it here, but only note that a corresponding isospectral pair
is constructed similarly to the case of the $n=2$. Namely, the
scaling, $\kappa_3>\kappa_2>\kappa_1$, and translation, $\tau_1$,
$\tau_2$ and $\tau_3$, parameters of the subsystem $H_3$ are taken
arbitrarily, the scaling parameters of the partner system $H'_3$ are
the same, and parameter $\tau'_3$ can take any finite value
restricted by the condition $\tau'_3\neq \tau_3$. The relation
$\mathcal{C}_2=\mathcal{C}_3$ defines $\tau'_2$ uniquely in terms of
the already chosen parameters, and then the equality
$\mathcal{C}_1=\mathcal{C}_2$ fixes uniquely the remaining
displacement parameter $\tau'_1$.

\section{Spin-$1/2$ particle interpretation}\label{spin12}

In this section, following ref. \cite{CJNP}, we discuss shortly a
spin-$1/2$ particle interpretation of the studied class of the
soliton systems (\ref{Hspin_n}), (\ref{V+V-n}). This,
particularly, will shed a new light on a peculiarity of the
special family of isospectral $n$-soliton systems characterized by
the first order supercharges.

Consider a non-relativist particle (electron) of mass
$m=\frac{1}{2}$, charge $e=-1$ and gyromagnetic ratio $g=2$
confined to a plane in the presence of electric field described by
a scalar potential $\phi(x,y)$ and perpendicular magnetic field
$B_z(x,y)$. The system is described by the Pauli Hamiltonian
\begin{eqnarray}\label{HPauli}
    &H=(-i\frac{d}{dx}+A_x)^2+
    (-i\frac{d}{dy}+A_y)^2 +
    \sigma_3B_z-\phi\,.&
\end{eqnarray}
Let us assume that electric and magnetic fields are homogeneous in
the direction $y$, $\phi=\phi(x)$, $B_z=B_z(x)$, and choose
$A_x=0$, $A_y=a(x)$. Then $B_z=\frac{d a}{dx}$, and the spinor
wave function can  be taken in the form
$\Psi(x,y)=e^{iky}\psi(x)$. The action of the Hamiltonian
(\ref{HPauli}) on a spinor $\psi(x)$ reduces to the matrix
Hamiltonian of the form (\ref{Hspin_n}) with
$V_\pm(x)=(k+a(x))^2-\phi \pm \frac{d a}{dx}$. Our system
(\ref{Hspin_n}), (\ref{V+V-n}) corresponds to the scalar electric
potential and magnetic field of a special form
\begin{equation}\label{el-mag-fields}
    \phi(x)=(a(x)+k)^2-\frac{1}{2}(V_n+V_n'),\quad
    B_z(x)=\frac{da}{dx}=\frac{1}{2}(V_n-V'_n)\,,
\end{equation}
given by the $n$-soliton, reflectionless  potentials $V_n$ and
$V'_n$.
 Taking into account Eq.
(\ref{lnW}), the potentials $\phi(x)$ and $a(x)$ can be written  in
terms of the corresponding Wronskians as
\begin{eqnarray}\label{el-mag-Wr}
    &\phi(x)=(a(x)+k)^2+\frac{d^2}{dx^2}
    \ln \left({W_n}{W'_n}\right),\quad
    a(x)=\frac{d}{dx}\ln \left(\frac{W'_n}{W_n}\right)+c_0\,,&
\end{eqnarray}
where $c_0$ is an integration constant.  Therefore, a spin-$1/2$
particle in the plane subjected to homogeneous in $y$ direction
electric and magnetic fields of the special form
(\ref{el-mag-fields}) is described by an exotic supersymmetry that
was investigated and described in the previous sections.

Let us show now that the systems (\ref{Hspin_n}), (\ref{V+V-n})
constructed from the special isospectral pairs of the $n$-soliton
potentials, which are characterized by the first first order
supercharges (alongside with the supercharges of order $2n$ and
bosonic integrals $\mathcal{P}_{n,a}$, $a=1,2$, being differential
operators of order $2n+1$), correspond to a case of a zero
electric field, i. e. a constant scalar potential $\phi$. First,
consider a one-soliton case for which
$V_1=-2\,\text{sech}^2\kappa(x+\tau)$ and
$V'_1=-2\,\text{sech}^2\kappa(x+\tau')$. For it,  $W_1=\cosh
\kappa(x+\tau)$ and $W'_1=\cosh \kappa(x+\tau')$. Putting the
integration constant $c_0=\kappa\coth\kappa(\tau-\tau')-k$, we
obtain
\begin{equation}\label{aD}
    a(x)=-\Delta(x)-k,\qquad
    \Delta(x)=\kappa(\tanh\kappa(x+\tau)-\tanh\kappa(x+\tau')-
    \coth\kappa(\tau-\tau')),
\end{equation}
that, up to the constant term $-k$, coincides exactly with the
superpotential that appears in the first order intertwining operator
(\ref{X1breve}).  The trigonometric identity
\begin{equation}\label{triginden}
    1-\tanh \alpha \tanh \beta -\coth(\alpha-\beta)(\tanh
\alpha-\tanh\beta)=0
\end{equation}
gives then $\phi=\kappa^2\coth^2\kappa(\tau-\tau')$, that is a
square of the constant $\mathcal{C}$ defined  in Eq. (\ref{defC}).

In the same way, for the special $n=2$ case discussed in Section
\ref{n=2special}, we find $a(x)=W(x)-k$, where $W(x)$ is the
superpotential appearing in the first order intertwining operator
(\ref{XWsuper}), and the scalar electric potential reduces to the
square of the constant $\mathcal{C}_1=\kappa_1\coth
\kappa_1(\tau-\tau')$, $\phi=\mathcal{C}_1^2$. This picture with
disappearing electric field is also valid for special isospectral
$n$-soliton systems with $n>2$, which were briefly discussed in
the previous section.

It is interesting to note that electric field can also be
eliminated in the self-isospectral case of reflectionless
P\"oschl-Teller systems having $n>1$ bound states, that
corresponds to the pair of mutually shifted soliton potentials
$V_n=-n(n+1)\kappa^2\,\text{sech}^2\kappa(x+\tau)$ and
$V'_n=-n(n+1)\kappa^2\,\text{sech}^2\kappa(x+\tau')$ with $n>1$.
This, however, can be done by the price of changing the
gyromagnetic ratio $g=2$ corresponding to the Pauli Hamiltonian
(\ref{HPauli}), to the value $g_n=\sqrt{2n(n+1)}$.  Indeed,
changing the magnetic term in (\ref{HPauli}) for
$\frac{1}{2}g_n\sigma_3B_z$, analogous analysis with employing the
identity (\ref{triginden}) results in
$a(x)=-\frac{1}{2}g_n\Delta(x)-k$, where $\Delta(x)$ is the same
as in Eq. (\ref{aD}), and
$\phi=\frac{1}{2}n(n+1)\kappa^2\coth^2\kappa(\tau-\tau')$.
According to the discussion in Sections \ref{n=2exactiso} and
\ref{nisospectrality}, a matrix system (\ref{Hspin_n}),
(\ref{V+V-n}) with mutually shifted reflectionless P\"oschl-Teller
potentials is characterized by the pairs of the supercharges to be
differential operators of orders $n$ and $n+1$. This picture can
be contrasted with a nonlinear supersymmetric structure appearing
in the Landau problem for a charged spin-$1/2$ particle with
special values of the gyromagnetic ratio $g=2n$, see ref.
\cite{KlPlmag}, where supersymmetry is generated by a pair of the
supercharges to be differential operators of order $n$.

\section{Discussion and outlook}

A generic supersymmetric quantum mechanical system with a $2\times
2$ matrix Hamiltonian, whose components are intertwined either by
first order Darboux  or higher order Crum-Darboux differential
operators, is described by two fermionic supercharges constructed
from the intertwining generators. The supercharges together with
the matrix Hamiltonian generate, respectively,  either linear or
nonlinear $N=2$ superalgebra.
For the linear supersymmetry (in the
sense of superalgebra), the system has one non-degenerate zero
energy level corresponding to the ground state in the case of the
non-broken supersymmetry, or only degenerate energy levels if the
supersymmetry is broken. For nonlinear supersymmetry case the
picture is more complicated, and the system can possess $0\leq
\ell\leq n$ non-degenerate states if nonlinear supersymmetry is of
order $n$, see \cite{PlKlanom}, \cite{AndIof} and references
therein.

We studied a special class of reflectionless systems with
super-partners having the same number $n$ of discrete energy levels
in their spectra. Each of super-partner potentials describes an
$n$-soliton solution of a nonlinear KdV equation that depends on $n$
scaling and $n$ translation parameters,  and satisfies corresponding
higher stationary equation of the KdV hierarchy. Because of the
peculiar, soliton nature of the composite matrix Hamiltonians, their
supersymmetric structure, on the one hand, turns out to be more rich
in comparison with a generic case, and, on the other hand, it
experiences   essential changes depending on relation between the
two sets of $2n$ parameters that characterize the partner
$n$-soliton potentials.

It is worth to stress here that according to the terminology we
used, the complete isospectrality breaking for a pair of $n$-soliton
potentials $V_n=V_n(\kappa_1,\ldots,\kappa_n,\tau_1,\ldots,\tau_n)$
and $V'_n=V_n(\kappa'_1,\ldots,\kappa'_n,\tau'_1,\ldots,\tau'_n)$
means that $\kappa_j\neq \kappa'_{j'}$ for all $j,j'=1,\ldots, n$,
and so, the energies of their bound states, $E_j=-\kappa_j^2$ and
$E'_j=-\kappa'_j{}^2$, have no coincidence, i.e. the extended system
(\ref{Hspin_n}), (\ref{V+V-n}) in this case has $2n$ discrete
non-degenerate levels. At the same time, the lowest, zero energy
level at the bottom of the continuous part of the spectrum of the
extended system is doubly degenerate, while all the energy levels
with $E>0$ inside the continuous spectrum  are four-fold degenerate.

There are four supercharges in the system (\ref{Hspin_n}),
(\ref{V+V-n}), two of which are composed from intertwining
generators $X_{2k+1}$ and $X_{2k+1}^\dagger$ to be differential
operators of the odd order $2k+1\leq 2n+1$, while two other
fermionic integrals  are constructed from intertwining generators
$Y_{2l}$ and $Y_{2l}^\dagger$ of the even order $2l\leq 2n$, such
that in general case the total order, $\vert X_{2k+1}\vert +\vert
Y_{2l}\vert$,  of the basic irreducible intertwining operators
satisfies a relation $2n+1\leq (2k+1)+2l\leq 4n+1$. The system
also possesses two bosonic diagonal matrix integrals composed from
nontrivial Lax operators of the $n$-soliton subsystems, $Z_{2n+1}$
and $Z'_{2n+1}$, which are differential operators of order $2n+1$
being the Crum-Darboux dressed form of the free particle momentum
$p=-i\frac{d}{dx}$. Operator $Z_{2n+1}$ ($Z'_{2n+1}$) detects all
the physical non-degenerate states of the subsystem $H_n$ ($H'_n$)
by annihilating them.

 When the two sets of the scaling parameters
are completely different, we have a complete isospectrality
breaking, and the irreducible intertwining generators are of the
orders $2n+1$ and $2n$. In this case $X_{2n+1}$ and $Y_{2n}$
intertwine the partner Hamiltonians $H_n$ and $H'_n$  via a
virtual free particle system.
Operator $Y_{2n}$ detects all the
bound states of the $H'_n$ subsystem, by annihilating them, while
$X_{2n+1}$ makes the same job and, additionally, annihilates the
non-degenerate state of the zero energy at the bottom of the
continuous spectrum. The eigenstates of the $H'_n$ not annihilated
by these intertwining operators are transformed by them into the
corresponding eigenstates of the $H_n$. The operators
$X_{2n+1}^\dagger$ and $Y_{2n}^\dagger$ do the same with the
eigenstates of $H_n$.
 The anticommutator between the supercharges
of differential orders $2n+1$ and $2n$ generates the diagonal Lax
integral $\mathcal{P}_{n;1}=-i\text{diag}\,(Z_{2n+1},Z'_{2n+1})$
multiplied by the order $n$ polynomial of the matrix Hamiltonian.
Both bosonic integrals, $\mathcal{P}_{n;1}$ and
$\mathcal{P}_{n;2}=\sigma_3\mathcal{P}_{n;1}$, commute
nontrivially with the supercharges. The Hamiltonian
$\mathcal{H}_n$ of the system plays a role of the multiplicative
central charge of the nonlinear superalgebra, whose structure is
insensible to the translation parameters of the potentials.

In the simplest case of $n=1$, when the scaling parameters
$\kappa_1$ and $\kappa_1'$ of the partner potentials coincide, a
kind of a channel for a direct `tunneling' between the partners is
opened, the third order operator $X_3$ is substituted for the
operator $X_1$ of the first order, that intertwines $H_1$ and
$H'_1$ directly, without communication via the virtual free
particle system, and bosonic integral $\mathcal{P}_{1;1}$
transmutes  into the central charge of the superalgebra, whose
structure starts to depend on the `tunneling distance'
$\tau_1-\tau'_1$.
Operator $X_1$ transforms now all the physical
eigenstates of the $H'_1$ subsystem into the corresponding
eigenstates of the $H_1$. In the case $n>1$, each time when any
two discrete energy levels of the partner subsystems coincide, the
basic intertwining operators $X$ and $Y$ undergo a reduction,
decreasing their total differential order in two, and a dependence
on a relative translation parameter associated with a pair of
coinciding scaling parameters appears in the superalgebraic
structure. The details of restructuring of supersymmetry
generators depend  on whether the discrete energy levels of the
partners of the same or different ordinal  numbers do coincide. A
structure of supersymmetry also suffers abrupt changes in the
orders of the basic irreducible intertwining operators, leaving
invariant their total sum, when the coincidence of translation
parameters, associated with the coinciding scaling parameters,
happens. The supersymmetry also experiences a restructuring for
another kind of correlation,
$\kappa_j\coth\kappa_j(\tau_j-\tau'_j)=\kappa_{j'}\coth\kappa_{j'}
(\tau_{j'}-\tau'_{j'})$, $j\neq j'$, between the translation
parameters associated with the coinciding pairs of discrete energy
levels of the different ordinal numbers, $j\neq j'$.

Only in the case of the exact isospectrality of the partners, when
all their discrete energy levels coincide pairwise, and as a
consequence, their transmission scattering amplitudes also
coincide, the bosonic integral $\mathcal{P}_{n;1}$ transmutes into
the central charge of the superalgebra. In this case the total
order $2n+1$ of the two basic irreducible intertwining operators
$X$ and $Y$ coincides with the differential order of bosonic
integrals. A particular case of such a situation corresponds to a
self-isospectral pair of P\"oschl-Teller systems.

{}From the viewpoint of supersymmetric structure we investigated,
the self-isospectral P\"oschl-Teller pairs possess, however, no
special properties when $n>1$, though the special subfamily of the
extended systems with exact isospectrality that we detected
corresponds to a generalization of the $n=1$ self-isospectral
case. For $n>1$, those special isospectral pairs with the scaling
and translation parameters correlated by means of $(n-1)$
equalities
$\kappa_1\coth\kappa_1(\tau_1-\tau'_1)=\kappa_{j}\coth\kappa_{j}
(\tau_{j}-\tau'_{j})$, $j=2,\ldots,n$, are described by the basic
irreducible  intertwining generators $X_1$ and $Y_{2n}$. For
$n>1$, the corresponding isospectral partner potentials have a
form different from each other, and if one of them is chosen  to
be a reflectionless  P\"oschl-Teller potential with $n>1$ bound
states, an isospectral partner does not belong to the
P\"oschl-Teller hierarchy of potentials. More precisely, we
identified and investigated in detail supersymmetric structure of
such a special pair in the case $n=2$, while we provided here only
general indications that the same happens for $n>2$. The special
family of the completely isospectral pairs of $n$-soliton systems
with $n>2$ requires a separate consideration and will be presented
elsewhere. The property $\vert X_1\vert=1$ means that any of the
two hermitian supercharges composed from the irreducible
intertwining generators $X_1$ and $X_1^\dagger$ may be identified
as a first order, Dirac type, Bogoliubov-de Gennes finite-gap
Hamiltonian that belongs to the AKNS integrable hierarchy.
{}From
another perspective, we also observed the peculiarity of the
special family of completely isospectral pairs with $\vert
X_1\vert=1$ from the viewpoint of interpretation of the matrix
Hamiltonian (\ref{Hspin_n}), (\ref{V+V-n}) in terms of the
non-relativistic spin-$1/2$ particle system. In this context, we
showed that all the family of self-isospectral reflectionless
P\"oschl-Teller systems also is special.

Analyzing  the changes of supersymmetric structure associated with
a coincidence of the scaling parameters, or, that is the same, of
the bound states energies, we referred to the opening of tunneling
channels conventionally. This might correspond nevertheless to
real tunneling processes in some applications of the exotic
supersymmetry, particularly, related to instantons.

We discussed the exotic supersymmetric structure from the standpoint
of a usual Shr\"odinger equation that corresponds to a potential
problem for a particle with a constant mass. It would be interesting
to reinterpret the results from a perspective of a quantum problem
for a particle with a position-dependent mass \cite{PositionMass}
having in mind possible applications for condensed matter physics.

As it was  noted, by displacing  the coordinate $x$ for a pure
imaginary constant, $x\rightarrow x+i\delta$, our analysis can be
generalized for the case of  $\mathcal{PT}$-symmetric quantum
systems \cite{PTrev}. Such a generalization seems to deserve a
special attention as it was proved to be useful for a particular
case of supersymmetric extensions of reflectionless P\"oschl-Teller
and related systems, that helped recently to clarify some
peculiarities in  the $\mathcal{PT}$-symmetric quantum mechanics
\cite{PT1}. Particularly, $\mathcal{PT}$-symmetric generalization
might be useful for applications in quantum optics.

As we mentioned,  $n=1$  and $n=2$  reflectionless P\"oschl-Teller
systems control the stability of the kink solutions in the
sine-Gordon, $\varphi^4$, and other exotic (1+1)-dimensional field
theoretical models~\footnote{Reflectionless $n$-soliton potentials
of a general form like that analyzed  in Section \ref{generaln=2}
for $n=2$ also appear in stability equations for kink solutions in
certain (1+1)-dimensional nonlinear field models, see
\cite{AJM}.}. By considering the doublets of these fields with
equal or different masses \cite{MSTB,AloMat}, one could expect
that the studied supersymmetric structure may reveal itself
somehow at the level of the symmetries of the corresponding kink
solutions.

We investigated exotic supersymmetry of soliton systems with the
primary focus on its quantum mechanical aspects. The intriguing
open  question is whether it can be related somehow to a
space-time symmetry of relativistic field systems having
topological solitons. The developments in the Section IV of
\cite{AJM} seems to point towards a positive answer to this
conjecture.

We discussed supersymmetric structure by choosing the diagonal Pauli
matrix as a grading operator $\Gamma$. Alternative choices for
$\Gamma$ related to reflection operators are also possible. They
provide the identification of the nontrivial integrals of motion as
fermionic and bosonic generators in a way different from that
described here. Particularly, the treatment of $\mathcal{P}_{n;a}$
as odd supercharges is possible, see
\cite{PlyANie,PlKlanom,PlyNie,CJPjpa,AraPly,MP1}. Supersymmetric
structures for alternative choices of $\Gamma$ can be computed by
employing the product relations of the intertwining generators and
Lax operators collected in Appendix. The alternative choices were
useful for identification of the hidden supersymmetric structure in
the systems described by the first order Bogoliubov-de Gennes
Hamiltonian, particularly, in those associated with the
Schr\"odinger $n=1$ isospectral pair considered here
\cite{PlyNie,PlyANie}. In this direction, it seems to be interesting
to apply the results on a special case of the two-soliton pairs with
exact isospectrality studied in Section 6.7 to the physics related
to the Gross-Neveu model.

Finally, it would be interesting to generalize our analysis for
finite-gap periodic systems, which also find many interesting
applications in physics \cite{CDP,PlyANie,Jak,Thies,BasDun1}. In
that case it seems to be natural to restrict the considerations to
the isospectral pairs.

\vskip0.2cm
 \noindent \textbf{Acknowledgements}

 The work has been partially
 supported by FONDECYT Grants 1095027 and 1130017,
 and by CONICYT (Chile), by DICYT (USACH), and by
 Spanish Ministerio de Educacion y Ciencia (DGICYT) under Grant
 FIS2009-10546.
 MP and JMG are grateful, respectively,
 to Universidad de Salamanca  and Universidad de Santiago de Chile
for hospitality, and to Benasque Center of Science for a stimulating
environment.

\section*{Appendix}\label{ap1}
\renewcommand{\theequation}{A.\arabic{equation}}
\setcounter{equation}{0}

Here we collect the products of the intertwining operators and Lax
operators to be necessary for computing the concrete superalgebraic
relations.

In the $n=1$  non-isospectral  case,  $\kappa_1\neq\kappa'_1$, the
basic products of intertwining operators  and Lax integrals are
\begin{eqnarray}\label{n=1nonXX01}
    &X_3X_3^\dagger=H_1(H_1+\kappa_1^2)(H_1+\kappa'^2_1),\qquad
    X_3^\dagger X_3=H'_1(H'_1+\kappa_1^2)(H'_1+\kappa'^2_1),&\\
    &Y_2Y_2^\dagger=(H_1+\kappa_1^2)(H_1+\kappa'^2_1),\qquad
    Y_2^\dagger Y_2=(H'_1+\kappa_1^2)(H'_1+\kappa'^2_1),&\label{n=1nonXX02}\\
    &X_3Y^\dagger_2=-Y_2X^\dagger_3=(H_1+\kappa'^2_1)Z_3,\qquad
    Y^\dagger_2X_3=-X^\dagger_3Y_2=(H'_1+\kappa^2_1)Z'_3,&\label{n=1nonXX03}\\
    &Z_3X_3=-H_1(H_1+\kappa_1^2)Y_2,\qquad
    X_3Z'_3=-H_1(H_1+\kappa'^2_1)Y_2,&\label{n=1nonXX04}\\
    &Z_3Y_2=(H_1+\kappa_1^2)X_3,\qquad
    Y_2Z'_3=(H_1+\kappa'^2_1)X_3\,,&\label{n=1nonXX05}\\
    &Z_3Z^\dagger_3=-Z_3^2=H_1(H_1+\kappa_1^2)^2,\qquad
    Z'_3Z'^\dagger_3=-Z'^2_3=H'_1(H'_1+\kappa'^2_1)^2.&\label{n=1nonXX06}
\end{eqnarray}
The products $X_3^\dagger Z_3$, $Z_3'X_3^\dagger$, $Y_2^\dagger Z_3$
and $Z'_3Y_2^\dagger$ are obtained by Hermitian conjugation of
(\ref{n=1nonXX04}) and (\ref{n=1nonXX05}). They are given by
expressions of the same form but multiplied by $-1$ because of the
property $Z_3^\dagger=-Z_3$, and with substitutions $H_1\rightarrow
H_1'$, $X_3\rightarrow X_3^\dagger$ and $Y_2\rightarrow
Y_2^\dagger$. Relations (\ref{n=1nonXX06}) are needed for computing
of the superalgebraic structures in the case of alternative choices
of the grading operator.

In the $n=1$ isospectral case $\kappa_1=\kappa'_1$, because of
reduction (\ref{Xbr}), some relations are changed for
\begin{eqnarray}\label{n1isoXX01}
    &\breve{X}_1\breve{X}^\dagger_1=H_1+\mathcal{C}^2,\quad
    \breve{X}_1^\dagger\breve{X}_1=H'_1+\mathcal{C}^2,&\\
    &\breve{X}_1Y^\dagger_2=Z_3+\mathcal{C}(H_1+\kappa_1^2),\quad
    Y_2\breve{X}^\dagger_1=-Z_3+\mathcal{C}(H_1+
    \kappa_1^2),&\label{n1isoXX02}\\
    &Z_3\breve{X}_1=\breve{X}_1Z'_3=\mathcal{C}(H_1+\kappa_1^2)
    \breve{X}_1-(H_1+\mathcal{C}^2)Y_2\,,&\label{n1isoXX03}\\
    &Z_3Y_2=Y_2Z'_3=(H_1+\kappa_1^2)((H_1+
    \kappa_1^2)\breve{X}_{1}-\mathcal{C}
    Y_2)\,.&\label{n1isoXX04}
\end{eqnarray}
The products $\breve{X}_1^\dagger Z_3=Z'_3\breve{X}_1^\dagger$ and
$Y_2^\dagger Z_3=Z'_3 Y_2^\dagger$ are obtained by Hermtian
conjugation of (\ref{n1isoXX03}) and (\ref{n1isoXX04}) as in the
non-isospectral case.

For a pair of $n$-soliton systems with complete isospectrality
breaking the basic products are
\begin{equation}\label{XXYYnnon01}
    Y_{2n}Y_{2n}^\dagger=\P_n \P_n'\,,
    \qquad
    X_{2n+1}X_{2n+1}^\dagger=H_n\P_n\P_n'\,,
\end{equation}
\vskip-0.9cm
\begin{eqnarray}
\label{XXYYnnon02}
   & X_{2n+1}Y_{2n}^\dagger=-Y_{2n}X_{2n+1}^\dagger=
    \P_n' Z_{2n+1}\,,&\\
\label{XXYYnnon03}
    &Z_{2n+1}Y_{2n}=\P_n X_{2n+1}\,,
    \qquad
    Y_{2n}Z{}'_{2n+1}=\P_n' X_{2n+1}\,,&\\
\label{XXYYnnon04}
   & Z_{2n+1}X_{2n+1}=-H_n\P_n Y_{2n}\,,
    \qquad
    X_{2n+1}Z{}'_{2n+1}=-H_n \P_n' Y_{2n}\,,&\\
\label{XXYYnnon05}
    &Z_{2n+1}^2=-H_n\P_n \,,&
\end{eqnarray}
where $\P_n=\prod_{l=1}^{n}(H_n+\kappa_l^2)$, $\P'_n=
\prod_{l=1}^{n}(H_n+\kappa_l'^2)$. Other products of the type
$X_{2n+1}^\dagger Y_{2n}$ etc. are obtained from these ones via
the change $\kappa_j\leftrightarrow \kappa'_j$,
$\tau_j\leftrightarrow \tau'_j$ with taking into account that
$X_{2n+1}^\dagger=-X'_{2n+1}$, $Y_{2n}^\dagger=Y'_{2n}$ and
$Z_{2n+1}^\dagger=-Z_{2n+1}$.

For three cases (\ref{break1}), (\ref{break2}), and (\ref{break2})
of $n=2$ pairs with partial isospectrality breaking, the basic
product relations are obtained from
(\ref{XXYYnnon01})--(\ref{XXYYnnon05}) by taking into account the
reduction relations (\ref{X5redX3A}), (\ref{X5redX3B}) and
(\ref{X5redX3AB}). The latter  are presented in the unified form
$X_5=h_{\kappa_i}\breve{X}^l_3-\mathcal{C}_lY_4$, and then for each
of three cases, distinguished by the index $l=1,2,3$ for
(\ref{break1}), (\ref{break2}), (\ref{break3}), respectively, we
have
\begin{eqnarray}\label{n2partial}
    &\breve{X}^l_3\breve{X}^{l\dagger}_3=
    h_{\mathcal{C}_l}h_{\kappa_d}h_{\kappa'_d}\,,\quad
    Y_4Y_4^\dagger=h^2_{\kappa_i}h_{\kappa_d}h_{\kappa'_d},&\\
    &\breve{X}^l_3Y_4^\dagger=h_{\kappa'_d}(Z_5+\mathcal{C}_l
    h_{\kappa_i}h_{\kappa_d}),\qquad
    Y_4\breve{X}^{l\dagger}_3=h_{\kappa'_d}(-Z_5+\mathcal{C}_l
    h_{\kappa_i}h_{\kappa_d}),&\\
    &Z_5Y_4=h_{\kappa_i}h_{\kappa_d}(h_{\kappa_i}
    \breve{X}^l_3-\mathcal{C}_lY_4),\quad
    Y_4Z'_5=h_{\kappa_i}h_{\kappa'_d}(h_{\kappa_i}
    \breve{X}^l_3-\mathcal{C}_lY_4),&\\
    &Z_5\breve{X}^l_3=h_{\kappa_d}(\mathcal{C}_lh_{\kappa_i}
    \breve{X}^l_3-h_{\mathcal{C}_l}Y_4),\quad
    \breve{X}^l_3Z'_5=h_{\kappa'_d}(\mathcal{C}_lh_{\kappa_i}
    \breve{X}^l_3-h_{\mathcal{C}_l}Y_4),&
\end{eqnarray}
where  $h_\alpha=H_2+\alpha^2$,
$\alpha=\kappa_i,\,\kappa_d,\,\kappa'_d,\,\mathcal{C}_l$.

The $n=2$ partial isospectrality breaking case
$\kappa_1=\kappa'_1$, $\kappa_2\neq \kappa'_2$, $\tau_1=\tau'_1$,
shown on Fig. \ref{fig4}a, is characterized by the following basic
products of the intertwining and Lax operators,
\begin{eqnarray}
    &\breve{Y}^A_2\breve{Y}^{A\dag}_2= h_{\kappa_2}
    h_{\kappa'_2}\,,\qquad
    X_5X^\dag_5=H_2h^2_{\kappa_1}
    h_{\kappa_2}h_{\kappa'_2}\,,&\label{AAA01}\\
    &X_5\breve{Y}^{A\dag}_2= -
    \breve{Y}^A_2X^\dag_5=h_{k'_2}Z_5\,,\qquad
    \breve{Y}^{A\dag}_2X_5=-X^\dag_5
    \breve{Y}^A_2=h_{\kappa'_2}Z'_5\,,&\label{AAA02}\\
    &Z_5X_5=-H_2h^2_{\kappa_1} h_{\kappa_2} \breve{Y}^A_2\,,
    \qquad
    X_5Z'_5=-H_2h^2_{\kappa_1}h_{\kappa'_2}
    \breve{Y}^A_2\,,\qquad&\label{AAA03}\\
    &Z_5\breve{Y}^A_2=h_{\kappa_2}X_5\,,\qquad
    \breve{Y}^A_2Z'_5=h_{\kappa'_2}X_5\,.& \label{AAA04}
\end{eqnarray}

For the $n=2$ isospectral case with a common $n=1$ virtual system,
when $\kappa_1=\kappa'_1$, $\kappa_2=\kappa'_2$, $\tau_1=\tau'_1$,
$\tau_2\neq \tau'_2$, the basic products are
\begin{equation}\label{n=2iso1101}
    \breve{Y}_2^A\breve{Y}_2^{A\dagger}=h_{\kappa_2}^2,\quad
    \breve{Y}_2^{A\dagger}\breve{Y}_2^A=h'^2_{\kappa_2},\quad
    \breve{X}_3^B\breve{X}_3^{B\dagger}=h_{\mathcal{C}_2}h_{\kappa_1}^2,\quad
    \breve{X}_3^{B\dagger}\breve{X}_3^B=h'_{\mathcal{C}_2}h'^2_{\kappa_1},
\end{equation}
\begin{equation}\label{n=2iso1102}
    \breve{X}_3^B\breve{Y}_2^{A\dagger}=
    Z_5+\mathcal{C}_2h_{\kappa_1}h_{\kappa_2},\quad
    \breve{Y}_2^{A}\breve{X}_3^{B\dagger}=
    -Z_5+\mathcal{C}_2 h_{\kappa_1}h_{\kappa_2},
\end{equation}
\begin{equation}\label{n=2iso1103}
    \breve{X}_3^{B\dagger}\breve{Y}_2^A=
    -Z'_5+\mathcal{C}_2 h'_{\kappa_1}h'_{\kappa_2},\quad
    \breve{Y}_2^{A\dagger}\breve{X}_3^B=
    Z'_5+\mathcal{C}_2 h'_{\kappa_1}h'_{\kappa_2},
\end{equation}
\begin{equation}\label{n=2iso1104}
    Z_5\breve{Y}_2^A=\breve{Y}_2^AZ'_5=
    h_{\kappa_2}^2\breve{X}_3^B-
    \mathcal{C}_2h_{\kappa_1}h_{\kappa_2}\breve{Y}_2^A,\quad
    \breve{Y}_2^{A\dagger}Z_5=Z'_5\breve{Y}_2^{A\dagger}=
    \mathcal{C}_2 h'_{\kappa_1}h'_{\kappa_2}
    \breve{Y}_2^{A\dagger} -
    h'^2_{\kappa_2}\breve{X}_3^{B\dagger},
\end{equation}
\begin{equation}\label{n=2iso1105}
    Z_5\breve{X}_3^B=\breve{X}_3^BZ'_5=
    \mathcal{C}_2h_{\kappa_1}h_{\kappa_2}
    \breve{X}_3^B-h_{\mathcal{C}_2}h_{\kappa_1}^2
    \breve{Y}_2^A,\,\,
    \breve{X}_3^{B\dagger}Z_5=Z'_5\breve{X}_3^{B\dagger}=
    h'_{\mathcal{C}_2}h'^2_{\kappa_1}
    \breve{Y}_2^{A\dagger}-\mathcal{C}_2h'_{\kappa_1}h'_{\kappa_2}
    \breve{X}_3^{B\dagger}.
\end{equation}
Here $h_{\kappa_i}=H_2+\kappa_i^2$, $h'_{\kappa_i}=H'_2+\kappa_i^2$
$i=1,2$, $h_{\mathcal{C}_2}=H_2+\mathcal{C}_2^2$,
$h'_{\mathcal{C}_2}=H'_2+\mathcal{C}_2^2$, and
$\mathcal{C}_2=\kappa_2\coth\kappa_2(\tau_2-\tau'_2)$. Relations for
the same isospectral case but with $\tau_2=\tau'_2$, $\tau_1\neq
\tau'_1$ are obtained from these ones by interchanging
$A\leftrightarrow B$, $\kappa_1\leftrightarrow \kappa_2$,
$\tau_1\leftrightarrow \tau_2$, $\tau'_1\leftrightarrow \tau'_2$ and
by, correspondingly,  changing $\mathcal{C}_2\rightarrow
\mathcal{C}_1$.

In generic $n=2$ isospectral case, $\kappa_1=\kappa'_1$,
$\kappa_2=\kappa_2'$, $\tau_1\neq \tau'_1$, $\tau_2\neq \tau'_2$,
denoting $\breve{X}_3^{(1)}=\breve{X}^A_3$ and
$\breve{X}_3^{(2)}=\breve{X}^B_3$, we have
\begin{eqnarray}\label{XXABXX}
    &\breve{X}_3^{(i)}\breve{X}_3^{(j)\dagger}=
    h_ih_jh_{ij}-(\mathcal{C}_i-\mathcal{C}_j)
    Z_5\,,\qquad
    \breve{X}_3^{(i)\dagger}\breve{X}_3^{(j)}=
    h'_ih'_jh'_{ij}+(\mathcal{C}_i-\mathcal{C}_j)
    Z'_5\,,&\\
\label{ZXABXZ}
    &Z_5\breve{X}_3^{(i)}=\breve{X}_3^{(i)}Z'_5=
    -\frac{1}{\Delta{C}}\left(
    (-1)^i h_1h_2h_{12}\breve{X}_3^{(i)}+\epsilon^{ij}h_{ii}h_j^2
    \breve{X}_3^{(j)}\right),&\\
\label{ZXABXZdag}
    &\breve{X}_3^{(i)\dagger}Z_5=Z'_5\breve{X}_3^{(i)\dagger}=
    \frac{1}{\Delta{C}}\left(
    (-1)^i
    h'_1h'_2h'_{12}\breve{X}_3^{(i)\dagger}+\epsilon^{ij}h'_{ii}h'^2_j
    \breve{X}_3^{(j)\dagger}\right),&
\end{eqnarray}
where $h_i=H_2+\kappa_i^2$, $h'_i=H'_2+\kappa_i^2$,
$h_{ij}=H_2+\mathcal{C}_i\mathcal{C}_j$,
$h'_{ij}=H'_2+\mathcal{C}_i\mathcal{C}_j$, $\Delta\mathcal{C}=
\mathcal{C}_2-\mathcal{C}_1$, and no summation in $i,j=1,2$ is
implied on the right hand sides.

For the $n=2$ special isospectral case
$\mathcal{C}_1=\mathcal{C}_2$,
\begin{eqnarray}\label{n2specialCC01}
    &\hat{X}_1\hat{X}_1^\dagger=h_{\mathcal{C}_1}\,,\qquad
    \hat{X}_1^\dagger\hat{X}_1^=h'_{\mathcal{C}_1}\,,\qquad
    Y_4Y_4^\dagger=h_{\kappa_1}^2h_{\kappa_2}^2\,,
    \qquad
     Y_4^\dagger Y_4=h'^2_{\kappa_1} h'^2_{\kappa_2}\,,&\\
    &\hat{X}_1Y_4^\dagger=Z_5+\mathcal{C}_1h_{\kappa_1}h_{\kappa_2}\,,\qquad
    Y_4\hat{X}_1^\dagger=-Z_5+\mathcal{C}_1h_{\kappa_1}h_{\kappa_2}\,,&
    \label{n2specialCC02}\\
    &Z_5\hat{X}_1=\hat{X}_1Z'_5=\mathcal{C}_1h_{\kappa_1}h_{\kappa_2}
    \hat{X}_1-h_{\mathcal{C}_1}Y_4\,,\quad
    \hat{X}_1^\dagger Z_5=Z'_5\hat{X}_1^\dagger=
    h'_{\mathcal{C}_1}Y_4^\dagger-\mathcal{C}_1h'_{\kappa_1}h'_{\kappa_2}
    \hat{X}_1^\dagger\,,&\label{n2specialCC03}\\
    &Y_4Z_5'=Z_5Y_4=h_{\kappa_1}h_{\kappa_2}
    (h_{\kappa_1}h_{\kappa_2}\hat{X}_1
    -\mathcal{C}_1Y_4)\,,
    \,\,\,
    Z_5'Y_4^\dagger=Y_4^\dagger Z_5=h'_{\kappa_1}h'_{\kappa_2}
    (\mathcal{C}_1Y_4^\dagger-h'_{\kappa_1}h'_{\kappa_2}
    \hat{X}_1^\dagger
    )\,,\qquad&\label{n2specialCC04}
\end{eqnarray}
where $h_{\kappa_i}=H_2+\kappa_i^2$, $h'_{\kappa_i}=H'_2+\kappa_i^2$
$i=1,2$, $h_{\mathcal{C}_1}=H_2+\mathcal{C}_1^2$,
$h'_{\mathcal{C}_1}=H'_2+\mathcal{C}_1^2$.

\end{document}